%% file: main.tex
\documentclass[10pt,twocolumn,letterpaper]{article}

\usepackage[pagenumbers]{cvpr} 

\input{preamble}

\definecolor{cvprblue}{rgb}{0.21,0.49,0.74}
\usepackage[pagebackref,breaklinks,colorlinks,allcolors=cvprblue]{hyperref}
\usepackage{xcolor}
\usepackage[table]{xcolor}
\usepackage{graphicx}
\usepackage{comment}
\usepackage{amsmath}
\usepackage{float}
\usepackage{placeins}
\usepackage{tabularx}
\usepackage[accsupp]{axessibility}

\def\paperID{****} 
\def\confName{CVPR}
\def\confYear{2026}

\title{MuCo: Multi-turn Contrastive Learning for Multimodal Embedding Model}

\author{Geonmo Gu$^{1,\,3}$ \quad Byeongho Heo$^{1}$ \quad Jaemyung Yu$^{1}$ \quad Jaehui Hwang$^{1}$ \quad Taekyung Kim$^{1}$ \\ \quad Sangmin Lee$^{3}$ \quad HeeJae Jun$^{2}$ \quad Yoohoon Kang$^{\ddagger,\,2}$ \quad Sangdoo Yun$^{\dagger,\,1}$ \quad Dongyoon Han$^{\dagger,\,1}$ \\
\\
{$^{1}${NAVER AI Lab} \qquad $^{2}${NAVER AI Search Platform} \qquad $^{3}${Korea University}} \\
}

\begin{document}

\maketitle
{\let\thefootnote\relax\footnotetext{\hspace{-1.5em}$^\ddagger$Provided general support for the project. $^\dagger$Corresponding authors.}}
\begin{abstract}
Universal Multimodal embedding models built on Multimodal Large Language Models (MLLMs) have traditionally employed contrastive learning, which aligns representations of query-target pairs across different modalities. 
Yet, despite its empirical success, they are primarily built on a ``single-turn'' formulation where each query-target pair is treated as an independent data point. This paradigm leads to computational inefficiency when scaling, as it requires a separate forward pass for each pair and overlooks potential contextual relationships between multiple queries that can relate to the same context.
In this work, we introduce Multi-Turn Contrastive Learning (\ours), a dialogue-inspired framework that revisits this process.
\ours\ leverages the conversational nature of MLLMs to process multiple, related query-target pairs associated with a single image within a single forward pass. 
This allows us to extract a set of multiple query and target embeddings simultaneously, conditioned on a shared context representation, amplifying the effective batch size and overall training efficiency.
Experiments exhibit \ours\ with a newly curated 5M multimodal multi-turn dataset (M3T), which yields state-of-the-art retrieval performance on MMEB and M-BEIR benchmarks, while markedly enhancing both training efficiency and representation coherence across modalities.
Code and M3T are available at \href{https://github.com/naver-ai/muco}{\texttt{https://github.com/naver-ai/muco}}
\end{abstract}


\section{Introduction}
\label{sec:intro}

\begin{figure*}[t]
\centering
\vspace{-0.5em}
\includegraphics[width=0.65\linewidth]{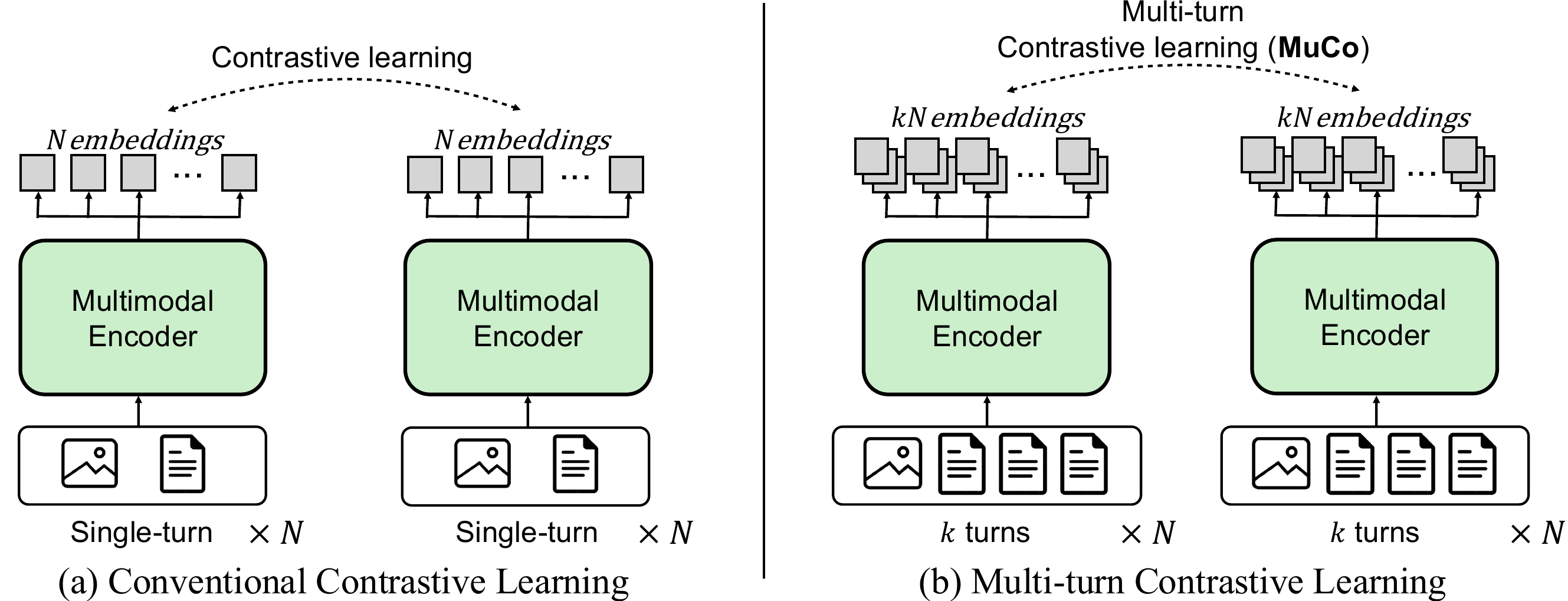}
\vspace{-0.5em}
\caption{\textbf{Comparison of conventional single-turn vs. our multi-turn contrastive learning.} (a) Conventional contrastive learning employs a single query-target pair per image, using negative targets from other images to learn discriminative representations. (b) Our multi-turn contrastive learning (\ours) generalizes this paradigm by using \textit{multiple} query-target pairs per image, with expanded negative targets corresponding to \textit{expanded} targets (from other images), which enables the model to learn more discriminative embeddings. 
Notably, for the same number of encoder forward passes, \ours\ provides $k$-times larger effective batch size than conventional contrastive learning.
}
\label{fig:simple_overview}
\vspace{-1.5em}
\end{figure*}

Universal multimodal embeddings aim to encode diverse modalities and tasks into a unified vector space, enabling a single model to generalize across a wide range of applications without task-specific adaptation. As real-world scenarios increasingly demand flexible and scalable multimodal understanding, Multimodal Large Language Models (MLLMs) have emerged as a powerful foundation for learning such universal representations. By jointly modeling visual and textual inputs, MLLM-based embedding frameworks have demonstrated remarkable performance across classification, Visual Question Answering (VQA), and multimodal retrieval tasks~\cite{jiang2025vlm2vec, gu2025unime, jiang2024e5v, liu2025lamra, lin2025mmembed, chen2025moca, chen2025mme5, yu2025cafe, faysse2025colpali}.

Despite this progress, most existing methods~\cite{radford2021learning, li2022blip, li2023blip2, gu2025unime, jiang2024e5v, liu2025lamra, lin2025mmembed, chen2025moca, chen2025mme5, yu2025cafe, faysse2025colpali} still rely on \textbf{``single-turn'' contrastive learning}~\cite{oord2018infonce}, which processes one query-target pair per image and relies on a large batch size to enhance their embedding quality. 
We argue that this paradigm has two critical limitations.
First, by treating each pair as an isolated data point, it overlooks the rich contextual interdependence among the potential multiple queries that can be derived from the same image. While a semantically rich image can support diverse text queries (\eg, object attributes, spatial relationships)~\cite{chunprobabilistic}, processing them independently may miss the opportunity to build a contextually coherent representation.
Second, scaling up contrastive learning by enlarging batch size introduces substantial computational overhead. This is because a large batch simply means processing more images. Unlike text, each image requires processing by a visual encoder and generates a substantially larger number of tokens. This inefficiency hinders the scalability of universal multimodal embedding models that depend on large-batch contrastive learning.

We introduce \textbf{Multi-turn Contrastive Learning} (\ours) – a novel method that extends traditional single-turn contrastive learning (\cref{fig:simple_overview}a) into a multi-turn setting (\cref{fig:simple_overview}b). Instead of processing isolated pairs, \ours\ processes multiple query-target pairs for a single image in a dialog format.
By leveraging the causal and conversational capability of MLLMs, \ours\ models the sequential query–target interactions across turns. This enables the model to progressively refine its embeddings through subsequent turns, yielding context-rich representations that generalize effectively across diverse tasks.
Furthermore, our method is efficient: the computationally heavy visual input is processed only once, while subsequent lightweight text-only turns are used to extract multiple distinct embeddings. This strategy dramatically increases the effective batch size with minimal additional training cost.

Our framework is strengthened through successive pretraining and fine-tuning stages. 
For pretraining, we construct a \textbf{5M}-scale multimodal multi-turn dataset featuring diverse multi-turn conversational pairs that explicitly model various tasks from a single image, synthesized via MLLMs~\cite{bai2025qwen2.5} and LLMs~\cite{agarwal2025gpt}. 
Pretraining on the M3T dataset allows \ours\ to benefit from both a large effective batch size and richer contextual signals from sequential turns, leading to a substantially more powerful and discriminative shared embedding space.
Subsequently, in the fine-tuning stage, we adopt the multi-turn paradigm to standard single-turn datasets.
This is achieved by simulating multi-turn reasoning via a prompt-based in-context reconstruction task (akin to masked modeling~\cite{devlin2019bert}), which further enhances the embedding's discriminative power. 

Extensive experiments demonstrate that \ours\ substantially outperforms existing MLLM-based embedding models across multiple benchmarks~\cite{meng2025vlm2vec2, wei2024uniir}, achieving state-of-the-art performance with improved efficiency and generalization.
\ours\ achieved new SOTA on the MMEB benchmark~\cite{jiang2025vlm2vec}, improving zero-shot performance (Precision@1) by +3.0\%p and fine-tuning performance by +1.6\%p over the previous SOTA~\cite{chen2025mme5,thirukovalluru2025b3}. On the M-BEIR benchmark~\cite{wei2024uniir}, \ours\ also secures SOTA in both 2B and 7B, improving overall Recall by +1.6\%p and +1.7\%p, respectively.
Notably, \ours\ demonstrates remarkable efficiency: while increasing the effective batch size 7 times ($1024\rightarrow7168$), the required FLOPs increase by a mere $\leq3\%$ (from 17.5 to 18.0 PFLOPs). This is substantially more efficient than standard contrastive learning, which requires 122.7 PFLOPs for the same batch size expansion, and achieves +0.7\%p higher performance.

\section{Related Work}
\label{sec:relatedwork}

\noindent\textbf{MLLM-based universal multimodal embedding model.}
Recent studies on universal multimodal embedding can be grouped into two main paradigms: CLIP-based methods~\cite{radford2021learning, li2022blip, li2023blip2, zhai2023sigmoid} introduced architectural or objective variations to improve visual-text alignment. In contrast, MLLM-based approaches
leveraged the strong multimodal understanding and instruction-following capabilities of large multimodal language models (MLLMs) to produce more semantically aligned and instruction-aware embeddings.
Within this paradigm, one line of work focused on scaling and curating data for embedding learning, such as VLM2Vec~\cite{jiang2025vlm2vec}, MMRet~\cite{zhou2024megapairs}, mmE5~\cite{chen2025mme5}.
A complementary line of research improved the training procedure itself.
E5-V~\cite{jiang2024e5v} unified representations through MLLM distillation, while UniME~\cite{gu2025unime} and LLaVE~\cite{lan2025llave} improved supervision through refined negative sampling.
B3~\cite{thirukovalluru2025b3} enhanced scalability via optimized batch construction, and M3Task-UEM~\cite{sharma2025multi} introduced task-adaptive representation learning to improve embedding effectiveness across diverse tasks.
While these models advance the field, they predominantly rely on single-turn supervision.

\noindent\textbf{Training data for the universal multimodal embeddings.}
Advances in data construction have strongly shaped universal multimodal embedding. Early web-crawled datasets like DataComp~\cite{gadre2023datacomp}, Conceptual Captions~\cite{sharma2018cc3m, changpinyo2021cc12m}, and YFCC~\cite{thomee2016yfcc100m} enabled large-scale alignment but suffered from noise and lacked semantic richness. The next generation~\cite{gu2024compodiff, zhang2024magiclens, zhou2024megapairs} introduced synthetic captioning for existing images~\cite{zhang2024magiclens, zhou2024megapairs} or synthesized images~\cite{gu2024compodiff}, improving linguistic diversity and grounding quality.
Later efforts focused on large-scale aggregation. MMEB benchmark~\cite{jiang2025vlm2vec} unified 36 datasets across four meta-tasks, and M-BEIR~\cite{wei2024uniir} consists of 16 retrieval datasets across 8 retrieval tasks. These two benchmarks are widely used to evaluate universal multimodal embeddings. More recently, mmE5~\cite{chen2025mme5} moved beyond aggregation by automatically generating aligned supervision, addressing data insufficiency, and extending to multilingual domains. Despite these advances, most data still relies on single query-target pairs. Recent MLLM-based embedding models focus on one-shot descriptions or isolated instructions, leaving their multi-turn conversational nature underexploited. To address this gap, our work explores how multi-turn, dialogue-inspired supervision can enhance contextual alignment and instruction grounding in universal multimodal embedding learning.

\section{Preliminary}
\label{sec:preliminary}

\noindent\textbf{Multimodal Large Language Models (MLLMs)}
\cite{liu2023llava, li2024llavanext, li2024llavaonevision, wang2024qwen2, bai2025qwen2.5} are generative models that extend beyond text-only inputs, processing a combination of images and text to produce sequential text tokens. Employing MLLMs as the backbone for multimodal embedding models is highly advantageous, as this enables the creation of unified embeddings from free-form multimodal samples. 
However, a fundamental challenge arises from the inherent asymmetry between modalities. 
The multimodal embedding models still rely on a visual encoder that converts each image into a dense sequence of tokens.
For instance, on the MMEB benchmark~\cite{jiang2025vlm2vec}, an image with the average resolution ($509\times456$, 294 tokens) requires 2.24 TFLOPs to process, while an average-length text (25 tokens) requires only 0.12 TFLOPs\footnote{Based on Qwen2-VL-2B~\cite{wang2024qwen2}}.
This token length imbalance makes the visual input far more computationally costly to process, creating a significant bottleneck for training efficiency.

\noindent\textbf{Contrastive learning.}
Previous methods~\cite{jiang2025vlm2vec, chen2025mme5, chen2025moca, gu2025unime, lin2025mmembed, zhou2024megapairs, thirukovalluru2025b3} for training multimodal embedding models typically rely on a contrastive learning objective, InfoNCE loss~\cite{oord2018infonce}, applied to single query-target pairs, where query and target consist of multimodal data such as images and texts: 
\begin{equation}
\label{eq:infonce}
\mathcal{L} = \frac{1}{|\mathcal{B}|} \sum_{(q_i, p_i) \in \mathcal{B}} -\log \frac{\phi(q_i, p_i)}{\sum\limits_{p \in \mathcal{N} \cup \{p_i\}} \phi(q_i, p)},
\end{equation} 
where $q_i$ is a query and $p_i$ is its positive target. The similarity score is given by $\phi(q_i, p_i)=\exp(f(q_i)^\top f(p_i)/\tau)$, where $f(\cdot)$ denotes an encoder producing $\ell_2$-normalized embeddings and $\tau$ is the temperature. $\mathcal{B}$ denotes the mini-batch, and $\mathcal{N}$ represents the set of valid negative targets within the batch.
While using \cref{eq:infonce} has become the de facto standard for contrastive learning, its effectiveness depends heavily on the batch size~\cite{radford2021learning, zhai2023sigmoid, chen2020simclr, chen2022we, gao2021scaling}, as larger batches provide more negative samples, which further leads to severe scalability issues in multimodal training.

\section{Method}
\label{sec:Method}

This section introduces our proposed method \textbf{Multi-turn Contrastive learning (\ours)}.
\ours\ capitalizes on the inherent causal processing capability of MLLMs, extending it to sequential multi-turn input processing such that each embedding (for each query) reflects the contextual dependencies of all prior turns. 
In effect, the model is compelled to internalize richer and coherent entangled embeddings that extend beyond conventional single-query alignment.

We implement \ours\ through a two-stage training process.
In the first pretraining stage (\S\ref{sec:muco_pretraining}), the model is trained on our newly constructed Multi-Modal Multi-Turn (\textbf{M3T}) dataset. This large-scale, dialogue-oriented corpus, containing multiple query-target pairs per image, enables the model to learn contextually coherent representations from conversational data.
Subsequently, the fine-tuning stage extends the proposed \ours\ framework on standard single-turn-based benchmarks, ensuring its effectiveness under conventional evaluation protocols (\S\ref{sec:muco_finetuning}).

\begin{figure*}[t]
    \centering
    \vspace{-0.5em}
    \includegraphics[width=.9\textwidth]{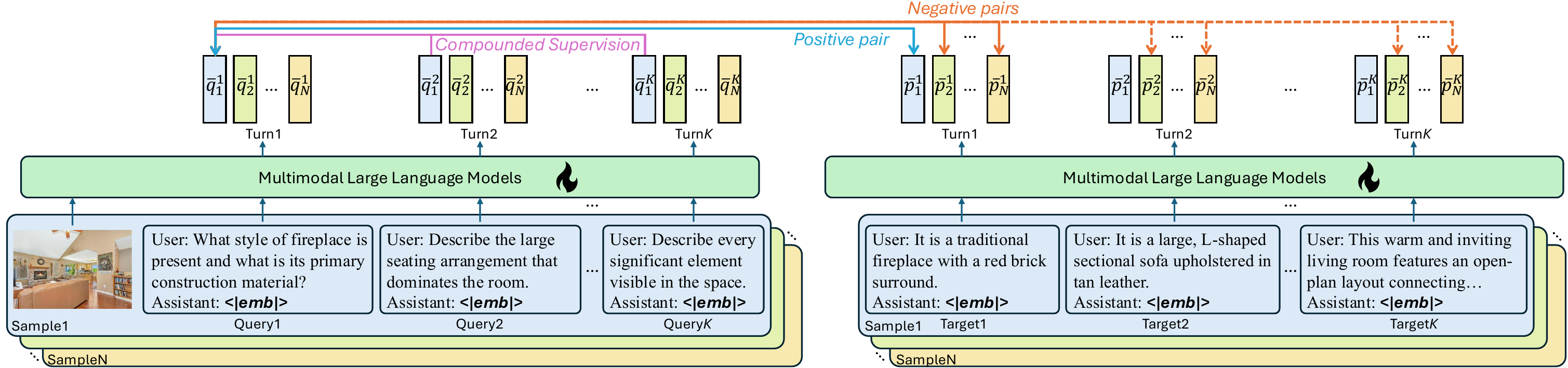}
    \vspace{-.5em}
    \caption{\textbf{Overview of \ours.}
    With a multiple query-target paired dataset, \ours\ intuitively structures the input by sequentially arranging the pairs as distinct dialogue turns.
    Lines are drawn between extracted embeddings, where \textbf{blue lines} denote a positive pair, \textbf{orange lines} denote negative pairs, and \textbf{purple lines} denote compounded supervision from subsequent turns to the initial turn. For clarity, lines are shown originating only from the earliest turn of the first sample in the batch. \textbf{Solid lines} represent the pair set used by conventional methods, while \textbf{dotted lines} represent the augmented pairs (\ie, more learning signals) contributed by 
    \ours\ framework. For visual clarity, we omit the embedding function notation (\eg $f(\cdot)$).
    }
    \label{fig:overview}
    \vspace{-1.5em}
\end{figure*}

\subsection{Multi-turn Contrastive Learning (\ours)}
\label{sec:muco_pretraining}
\label{sec:pretraining_dataset}

Our method reframes the embedding extraction process as a multi-turn dialogue. By placing a special embedding token (\texttt{<|emb|>}) in the response of the \texttt{assistant} within a prompt structure, as shown in \cref{fig:overview}, our framework can extract multiple distinct embeddings from a single input sequence at once.
Subsequent turns per image consist of appending \textit{text-only} queries or positive targets.

We first instantiate \ours\ in pretraining by compounding supervisory signals across sequential turns. Our goal is to learn context-rich embeddings for earlier turns -- particularly for the initial turn\footnote{We refer to the position of the first embedding token (\texttt{<|emb|>}) as the initial or earliest turn, which is only used for inference later.} -- that generalize effectively across tasks such as classification, retrieval, and visual question answering.
We believe this could be achieved by training on multiple query-target pairs, processed sequentially via our dialogue template (\cref{fig:overview}); this design ensures that earlier embeddings become effective through training subsequent turns, as they are continually referenced by later queries (\eg, the $n$-th embedding is formed after attending to the previous $n-1$ queries).

Let the set of training data be denoted as $\mathcal{D} = \{((\mathcal{I}_i, q_i^j), p_i^j)\}$ where $\mathcal{I}_i$ is an $i$-th image and $q_i^j, p_i^j$ represent $j$-th query text and positive target text. For a mini-batch $\mathcal{B} \subset \mathcal{D}$, employing the traditional contrastive learning loss \cref{eq:infonce} gives \vspace{-0.5em} 
\begin{equation}
    \frac{1}{|\mathcal{B}|} \sum_{((\mathcal{I}_i, q_i^j), p_i^j) \in \mathcal{B}} - \log \frac{\phi((\mathcal{I}_i, q_i^j), p_i^j)}{\sum\limits_{p \in \mathcal{N}\cup\{p_i^j\}} \phi((\mathcal{I}_i,q_i^j),p)}.
    \label{eq:naive_infonce}
\end{equation}
We believe this naive approach has two drawbacks. First, it often fails to leverage the full context of multiple query-target pairs for a given single image, as trained exclusively in a single-query setting, despite the fact that these pairs are not mutually exclusive (\ie, crucial cues may lie in other questions and answers). Second, when increasing the number of images by enlarging the batch size, it usually fails to mitigate the computational overhead incurred by the image parts.
To address these drawbacks, we propose concatenating queries and positive targets that share the same image using the prompt template shown in \cref{fig:overview}: 
\begin{equation}
    \bar{q}_i^j = (\mathcal{I}_i, (q_i^l)_{l \leq j}), \quad \bar{p}_i^j = (p_i^l)_{l \leq j}.
\end{equation}

\begin{figure}[t]
    \centering
    \vspace{-.5em}
    \includegraphics[width=.9\columnwidth]{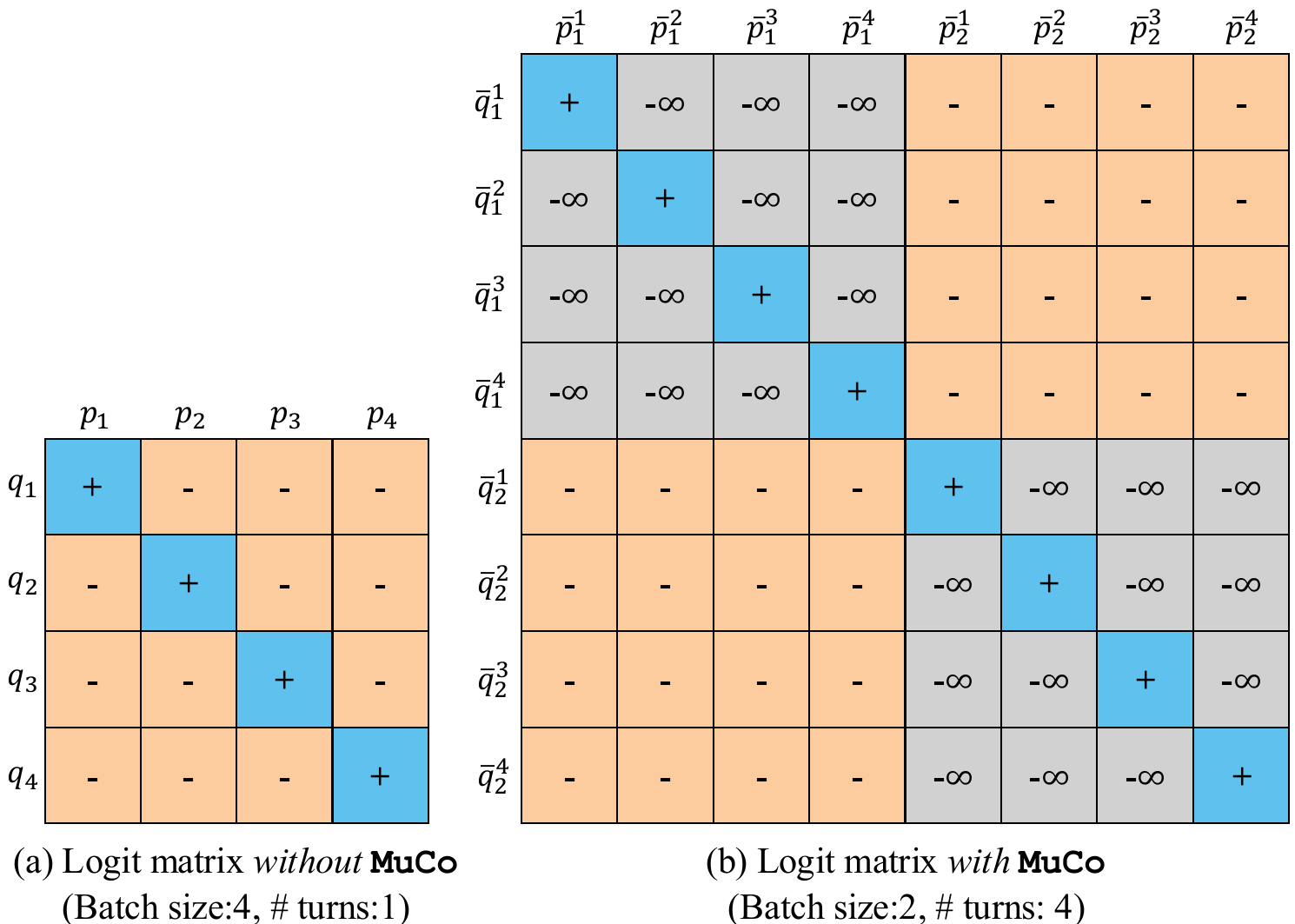}
    \vspace{-.5em}
    \caption{\textbf{Logit masking strategy in our \ours\ framework.}
    (a) The conventional method with a batch size of $N=4$ yields a $N \times N$ (\ie $4 \times 4$) matrix. In contrast, \ours\ (b) uses a batch size of $N=2$ and $k=4$ turns to construct a larger $Nk \times Nk$ (\ie $8 \times 8$) matrix. Crucially, our method masks out pairs originating from the same image (gray, $-\infty$) to prevent a semantic overlap issue. True positives (blue, $+$) and true negatives (orange, $-$) are used for the loss. Crucially, other pairs originating from the same image (gray, $-\infty$) are masked to prevent a semantic overlap issue.
    }
    \label{fig:cossim}
    \vspace{-1.5em}
\end{figure}

This cumulative structure enables processing multiple queries and positive targets in a single forward pass. By placing \texttt{<|emb|>} tokens at the end of each turn, we simultaneously extract all embeddings without re-encoding the image, significantly reducing computational overhead. Furthermore, gradients from all subsequent turns could flow back to the initial turn.
The loss function for multiple queries and targets is changed as
\begin{align}
\label{eq:muco_pretraining_infonce}
    \mathcal{L}_\textbf{\texttt{MuCo}}=&\frac{1}{|\mathcal{B}|} \sum_{(\bar{q}_i^j, \bar{p}_i^j) \in \mathcal{B}} - \log \frac{\phi(\bar{q}_i^j,\bar{p}_i^j )}{\sum\limits_{\bar{p}\in\mathcal{N}_i\cup\{\bar{p}_i^j\}} \phi(\bar{q}_i^j,\bar{p})},\\
\text{s.t.} \quad &\mathcal{N}_i =
\{\bar{p}_k^l\mid (\cdot, \bar{p}_k^l) \in \mathcal{B},\ k \neq i \}.
\end{align}
Note that to address the second drawback, the semantic overlap issue, we exclude all other positive targets originating from the same image $\mathcal{I}_i$ as the current query $\bar{q}_i^j$ from its negative set $\mathcal{N}_i$ (the denominator in \cref{eq:muco_pretraining_infonce}). This exclusion is practically achieved by adding negative infinity to the corresponding values in the logit matrix for the pairs to be masked, as shown in \cref{fig:cossim} (b).
This approach is better than \cref{eq:naive_infonce} in optimizing computation overheads.
For example, with a batch size of 1,024 images and 7 pairs per image, our method provides each query with 7,161 effective negative samples ($1024 \times 7 - 7$).

In this way, \ours\ inherently processes inputs as sequential turns, where each subsequent embedding is conditioned on the context of all preceding turns, enabling the model to learn diversified and context-aware relationships between multiple queries and a given image.
From another perspective, gradients from later turns retroactively refine earlier ones, yielding a compounded supervisory signal that encourages the initial representations to remain informative across the entire dialogue.

\noindent\textbf{Dialogue-templated data (M3T).} We construct synthetic corpora, named M3T,
to train \ours\ effectively, as such dialogue-oriented multimodal structures are publicly unavailable. Our key design principles are: we prioritize a large scale and multiple query-target pairs for an image for versatility and diversity, and structure the data to reflect comprehensive visual information, ensuring a thorough interpretation of both global and local image contexts. Furthermore, to maximize computational efficiency during training, we restrict positive targets to text-only formats, while images are exclusively used as part of the queries.

We take 5 million images sampled from DataComp~\cite{gadre2023datacomp}, processed through a two-step pipeline combining state-of-the-art MLLM~\cite{bai2025qwen2.5} and LLM~\cite{agarwal2025gpt}. First, we employ the MLLM to generate dense captions, distilling rich visual information into text. Second, we let the LLM synthesize seven diverse query-target pairs per image using the dense captions, aligned with MMEB core categories~\cite{jiang2025vlm2vec}: one for classification, one for retrieval, and five distinct VQA pairs composed of two requiring holistic understanding, two focusing on localized details, and one demanding creative reasoning. This process yields 35 million query-target text pairs and 5 million images. Further details and prompts are provided in the supplementary material.

\subsection{Single-turn Adaptation of \ours}
\label{sec:muco_finetuning}
Our multi-turn pretraining generalizes to the conventional multimodal benchmarks~\cite{jiang2025vlm2vec, wei2024uniir}, having a single query-target pair per sample.
To align with single-query data, we introduce an adaptive strategy that employs extra contextual dialogue from a given query, allowing \ours\ to elicit its pre-trained multi-turn capability.

\begin{figure}[t]
    \centering
    \vspace{-.5em}
    \includegraphics[width=.9\columnwidth]{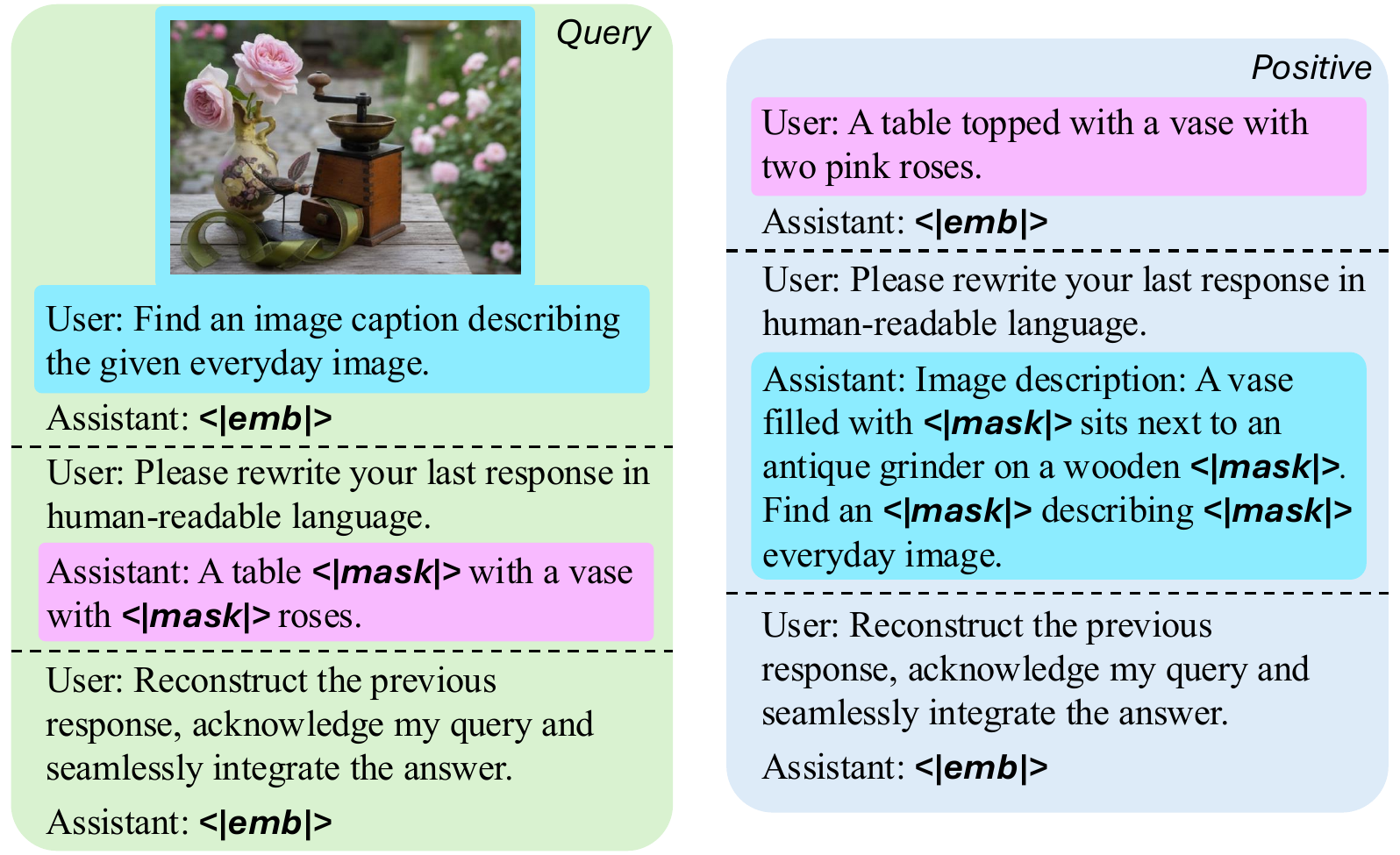}
    \vspace{-.5em}
    \caption{\textbf{Multi-turn template for fine-tuning \ours\ on single-pair datasets.}
    We illustrate Query (left) and Positive (right) templates. The initial query (cyan) is reused as a masked target on the Positive side, and the positive target (pink) becomes a masked target on the Query side. This process simulates multi-turn interactions from a single pair, guiding the model to reconstruct its counterpart and enrich the learned embeddings.
    }
    \label{fig:template_finetuning}
    \vspace{-1.5em}
\end{figure}

We repurpose \ours\ to handle single-query pairs.
As shown in \cref{fig:template_finetuning}, we first establish an initial turn to extract the initial embedding from the query $q$ or positive target $p$. In the following turns, we provide the model with its masked counterpart (\eg, the masked positive target for a given query). The model is then explicitly instructed to first reconstruct the masked parts and generate a subsequent embedding capturing the context of the entire dialogue.

We argue this process functions as an in-context reconstruction task, compelling the model to reason about the relationship between the query and target to infer the masked content.
We achieve this by leveraging LLMs' capabilities through prompting (\ie, instructing the LLM to perform reconstruction) and a special mask token (\texttt{<|mask|>}); interestingly, even simple tokens such as a space or `\_' work equally well. This prompt-only approach avoids auxiliary objectives, ensuring simplicity and direct application to standard contrastive learning.
To maintain the efficiency of text-only subsequent turns, we convert any images in the counterpart to text via image captioning. This conversion is performed offline, prior to finetuning, using a image captioning model. This approach offers flexibility, as any suitable captioning model can be employed for this preprocessing step (See Tab~\ref{tab:abl_finetuning_design}).
Crucially, we use only the initial embedding without the subsequent turn to align with standard single-query test scenarios~\cite{jiang2025vlm2vec, wei2024uniir}.

\input{tables/main_mmeb}
\input{tables/main_mbeir_small}

Details work as follows: first, the query $q$ and the corresponding positive target $p$ are augmented using prompts.
The augmented query and target $(q', p')$ is defined as 
$q' = (q, \pi_1, \Tilde{p}, \pi_2)$, $p' = (p, \pi_1, \Tilde{q}, \pi_2)$,
where $\pi_1$ is a describing request such as \textit{ User: Please rewrite your last response in human-readable language}, $\Tilde{p}$ and  $\Tilde{q}$ represents the masked query and target, and $\pi_2$ requests a new embedding such as  \textit{User: Reconstruct the previous response, acknowledge my query, and seamlessly integrate the answer}.

For original mini-batch $\mathcal{B}$, we utilizes augmented set $\bar{\mathcal{B}}$:
\begin{equation}
\label{eq:single_augmented_batch}
\bar{\mathcal{B}} = \bigcup_{(q,p)\in\mathcal{B}}
\{\, (q,p),\ (q,p'),\ (q',p),\ (q',p') \,\}.
\end{equation}
The contrastive loss with the augmented set is defined as
\begin{equation}
\label{eq:masked_infonce_single}
    \frac{1}{|\bar{\mathcal{B}}|} \sum_{(q_i, p_i) \in \bar{\mathcal{B}}} - \log \frac{\phi(q_i, p_i)}{\sum\limits_{p \in \mathcal{N}_{p_i}\cup\{p_i\}} \phi(q_i,p)}.
\end{equation}
Since the augmented samples are similar to the original ones, we exclude them from negative samples in contrastive loss. \ie, the negative set $\mathcal{N}_{p_i}$ is
\begin{equation}
    \mathcal{N}_{p_i} = \{\, p \mid (\cdot,p)\in\bar{\mathcal{B}}\,\}
    \setminus
    \{\, p_i,\ p_i^{-} \,\},
\end{equation}
where $p_i^{-}$ denotes the opposite form of $p_i$: it is the augmented version if $p_i$ is an original target in $\mathcal{B}$, and the original target if $p_i$ is an augmented one.
With this loss, the supervisory signal flows back to refine the initial embedding, enabling it to encapsulate richer relational information.
In essence, our key intuition is to use simple prompting (without any further complicated elements) to leverage the MLLM’s capability to simulate multi-turn behavior on single-turn data, thereby exploiting its pretraining strength. Simple tweaks such as token masking are also believed to enhance the discriminative power of the learned embeddings.
We examine several related design choices in the experimental section.

\section{Experiment}
\label{sec:exp}

\subsection{Experimental Setup}
\noindent\textbf{Implementation details.}
Qwen2-VL~\cite{wang2024qwen2} is utilized as the MLLM backbone. We apply LoRA~\cite{hu2022lora} with a rank of 64 and a scaling factor of $\alpha=64$ exclusively to the LLMs components of the model and freeze the visual encoder of the model. All training is conducted for 1 epoch on 32 NVIDIA A100 80GB GPUs. Unless otherwise specified, a global batch size of 1,024 is used. The contrastive learning temperature $\tau$ is set to $0.02$, and a constant learning rate of $5e^{-5}$ is employed.
For our pretraining (\S\ref{sec:muco_pretraining}), multiple query-target pairs are randomly shuffled per batch instance to prevent positional bias.
In our adaptive prompting strategy (\S\ref{sec:muco_finetuning}), we split the text sentence into words based on spaces and then randomly mask 50\% of the words using a uniform sampling strategy. If not specified, we employ Qwen2-VL-7B for image captioning in \S\ref{sec:muco_finetuning}

\noindent\textbf{Datasets.}
In \cref{tab:main_mmeb}, we use the proposed M3T dataset of 5M samples to pretrain \ours. For pre-training dataset analysis in \cref{tab:abl_pretraining}, we construct two randomly sampled subsets, M3T (20\%) with 1M and M3T (60\%) with 3M samples. We also utilize mmE5~\cite{chen2025mme5} dataset with 560K and the MegaPairs~\cite{zhou2024megapairs} dataset with 26M samples for comparison. The details of these datasets are described in the supplementary material. For the fine-tuning stage, models are trained on two benchmarks: MMEB~\cite{jiang2025vlm2vec} and M-BEIR~\cite{wei2024uniir}.
MMEB is a benchmark containing 36 individual datasets across four categories: classification, visual question answering (VQA), retrieval, and visual grounding. MMEB reframes all four categories into ranking problems with a maximum of 1,000 candidates. The benchmark provides training samples from 20 of the 36 datasets (considered In-Distribution), while the remaining 16 are reserved exclusively for evaluation (Out-of-Distribution).
M-BEIR is a benchmark focusing specifically on multimodal retrieval performance. It consists of 16 retrieval datasets across 8 retrieval tasks.
In our experiments, we fine-tune our models on the respective training set for each benchmark and evaluate them on the corresponding test set.

\subsection{Main Results}
We provide the evaluation results on MMEB in \cref{tab:main_mmeb} and M-BEIR in \cref{tab:main_mbeir}. \ours\ achieves state-of-the-art (SOTA) performance on the MMEB and M-BEIR benchmark across different model scales. This superior performance stems from our multi-turn dialogue structure, which allows the model to learn rich context and compounded supervisory signals, thereby effectively improving the embeddings. In the zero-shot setting, our 7B model (61.6) surpasses the previous SOTA mmE5-11B~\cite{chen2025mme5} (58.6). In the fine-tuning setting, \ours\ establishes new SOTA scores in both sub-7B and 7B-and-above categories. Specifically, \ours-2B achieves 69.5, outperforming the previous best B3-2B~\cite{thirukovalluru2025b3} (68.1). Similarly, \ours-7B establishes a new top score of 73.6, surpassing B3-7B~\cite{thirukovalluru2025b3} (72.0) and achieving the highest OOD score, which indicates superior generalization to unseen tasks.
This strong performance trend continues on the M-BEIR benchmark. \ours-2B (51.6) and \ours-7B (56.6) ourperform all competing methods. We note their robust performance in the complex multi-modal to multi-modal ($M \to M$) setting, achieving top scores of 64.8 (2B) and 70.4 (7B).

\subsection{Empirical Analysis}
This section analyzes our design choices. Unless specified, all experiments use \ours-2B on MMEB. In tables, \textit{ZS MMEB} denotes zero-shot performance (Precision@1) after \ours\ pretraining (\S\ref{sec:muco_pretraining}), and \textit{FT MMEB} denotes the performance (Precision@1) after \ours\ fine-tuning (\S\ref{sec:muco_finetuning}) on the MMEB training set.

\noindent\textbf{Impact of pretraining dataset.}
\input{tables/ablation_pretraining}
We analyze the impact of different pretraining datasets in \cref{tab:abl_pretraining} by varying pretraining method and dataset. Note that we fix \ours\ fine-tuning (\S\ref{sec:muco_finetuning}) to all cases for FT MMEB.
The results reveal two key insights. 
First, \ours\ pretraining is highly effective. Training on mmE5 after converting them to our multi-turn format outperforms the original single-turn pretraining (69.0 vs. 68.6 FT MMEB), demonstrating the significant advantage of our richer, multi-turn signal.
Second, M3T is a high-quality dataset that enables large-scale \ours\ pretraining with clear gains.
\ours\ pretraining performance improves monotonically from 0.6M to 5M. It implies that M3T offers mmE5-level quality at roughly $\times 10$ scale, while MegaPairs~\cite{zhou2024megapairs} degrades performance with massive data.
It is noteworthy that the model without any pretraining (\textit{None}) achieves 68.5, which is competitive with the mmE5-pretrained model (68.6). This shows that \ours\ fine-tuning alone is strong enough to surpass the previous SOTA B3-2B (68.1, Table \ref{tab:main_mmeb}) even without pre-training.

\input{tables/ablation_each_task}
We analyze each pretraining task's contribution in \cref{tab:abl_each_task}. Sequentially excluding tasks ($-$ CLS, global/local/creative VQA) causes a performance drop in the corresponding task. Notably, excluding local VQA also degrades Visual Grounding (GRD), and removing creative VQA further lowers overall performance. This confirms each component of our synthesized data is beneficial.

\noindent\textbf{Scaling batch-size vs. turns.}
\input{tables/ablation_multiturn_batchsize}
We compare scaling the batch size and the number of turns in \cref{tab:abl_multiturn_batchsize}. For the conventional method (\#turns=1), increasing the batch size from 1024 to 8192 yields performance gains (57.1 to 57.8) at a prohibitively high computational cost. In contrast, our method scales the number of turns (\#turns 2 to 7) at a fixed 1024 batch size. Notably, our 7-turn model (7168 effective batch size) achieves a higher score than the 1-turn baseline with a 7168 batch size (58.2 vs. 57.5) and even surpasses the larger 8192-batch baseline (57.8) while the 7168-batch baseline incurs a massive computation overhead.
This demonstrates that our method provides a far more efficient path to achieving the benefits of a larger effective batch.

\input{tables/ablation_causal_attention}
\noindent\textbf{Impact of compounded supervision.} We conduct an ablation study to validate the effectiveness of our compounded supervision (\ie, subsequent turns provide a compounded supervisory signal that retroactively refines earlier embeddings), as shown in \cref{tab:abl_causal_attention}.
We test a variant where this accumulation is disabled by modifying the causal attention mask in both pretraining and fine-tuning; this forces each turn to only attend to the initial image and its own tokens, isolating it from the context of preceding turns. The results demonstrate our hypothesis: disabling this compounded supervision leads to a performance drop (68.4) compared to \ours\ (69.5). This confirms that our strategy of accumulating supervisory signals across turns is a superior approach for learning robust representations.

\noindent\textbf{Impact of logit masking.}
\input{tables/ablation_intra_mask}
We validate the importance of our logit masking strategy in \cref{tab:abl_intra_mask}, which is critical for fine-tuning. Disabling it (\eg row (1,3) vs. row (2,4) in FT MMEB) causes a performance collapse. This is because, during fine-tuning, we construct subsequent turns using each counterpart. Without logit masking, the model treats these semantically overlapped pairs as negative pairs, leading to severe confusion that prevents learning. In contrast, the impact during pretraining is far less severe in ZS MMEB. This is because the pretraining data consists of diverse query-target pairs for each image. While this still causes a minor performance drop due to potential semantic overlap between the pairs, the diversity prevents the learning collapse seen in the fine-tuning stage.

\input{tables/ablation_finetuning_design}
\noindent\textbf{Ablation study of the subsequent turn design for fine-tuning on single-pair dataset.}
We compare other design choices for the subsequent turn used in the fine-tuning stage as shown in \cref{tab:abl_finetuning_design}. First, we analyze the counterpart masking ratio, finding that 50\% yields the best performance. A lower ratio (\eg 25\%) is suboptimal as the masked input is too similar to the original counterparts which provide only a minimal learning signal. Conversely, a higher ratio (\eg 75\%) also results in a larger performance drop because the reconstruction task becomes excessively difficult.

We also analyze the dialog template for the fine-tuning strategy.
Interestingly, using just a simple rephrasing template (\textit{User: Please rephrase your last response in embedding space\texttt{\symbol{92}n} Assistant: \texttt{<|emb|>}}) outperforms the \ours\ without compounded supervision (68.4 in \cref{tab:abl_causal_attention}). This demonstrates our compounded supervision can improve the performance even though a simple subsequent is employed.
When training without reconstruction guidance, the performance drops to 69.0 from 69.5, which confirms that the explicit reconstruction prompt is crucial for compelling the model to reason about the relationship between the pairs in Eq.~\ref{eq:single_augmented_batch}. Finally, we observe that the choice of image captioning model (BLIP-Large vs. Qwen2-VL-7B) has a minimal impact on performance.

\section{Conclusion}
In this work, we propose Multi-turn Contrastive Learning (\ours), a dialogue-inspired framework to overcome the limitations of conventional single-turn contrastive learning. By reframing representation learning as a multi-turn dialogue and modeling contextual dependencies, \ours\ learns richer and more coherent embeddings than isolated pair alignment.
Furthermore, \ours\ tackles scalability bottlenecks by processing multi-turn query sequences in a single forward pass. This approach significantly reduces computational overhead while increasing the effective batch size.
Supported by our new 5M-scale M3T corpus, \ours\ achieves new state-of-the-art performance on universal multimodal embedding benchmarks. \ours\ demonstrates that it is possible to simultaneously enhance model performance and training scalability, effectively redefining the efficiency-capacity trade-off in multimodal alignment. We believe this work opens new avenues for more efficient and context-aware multimodal representation learning systems.

\section*{Acknowledgment}
This work utilized Bruno from the NAVER AI Search Platform to facilitate large-scale data processing tasks. We thank the platform for providing robust Model as a Service (MaaS) support.

{
    \small
    \bibliographystyle{ieeenat_fullname}
    \bibliography{main}
}

\input{appendix_preprint}

\end{document}

%% file: tables/main_mmeb.tex
\definecolor{lightgray}{gray}{0.9}
\begin{table*}[t]
\centering
\caption{\textbf{Precision@1 (\%) results on MMEB}, which includes 36 tasks across four categories: Classification, Visual Question Answering (VQA), Retrieval, and Visual Grounding. ID and OOD stand for the in-distribution average and out-of-distribution average metrics, respectively.
\textbf{Boldface} denotes the best scores in the subset and the second-best scores are highlighted with \underline{underline}.}
\label{tab:main_mmeb}
\footnotesize
\vspace{-1em}
\begin{tabular}{lrcccc|cc|>{\columncolor{lightgray}}c}
\toprule
& & \multicolumn{4}{c}{\textbf{Per Meta-Task Score}} & \multicolumn{3}{c}{\textbf{Average Score}} \\
\cmidrule(lr){3-6} \cmidrule(lr){7-9}
\textbf{Models} & \textbf{\# Params} & \textbf{Classification} & \textbf{VQA} & \textbf{Retrieval} & \textbf{Grounding} & \textbf{ID} & \textbf{OOD} & \textbf{Overall} \\
\midrule
\multicolumn{9}{c}{\textit{Zeroshot setting (pretrained) on MMEB benchmark}} \\
\midrule
CLIP~\cite{radford2021learning} & 0.4B & 42.8 & 9.1 & 53.0 & 51.8 & -- & -- & 37.8 \\
MagicLens~\cite{zhang2024magiclens} & 0.6B & 38.8 & 8.3 & 35.4 & 26.0 & -- & -- & 27.8 \\
E5-V~\cite{jiang2024e5v} & 8B & 21.8 & 4.9 & 11.5 & 19.0 & -- & -- & 13.3 \\
MMRet~\cite{zhou2024megapairs} & 7B & 47.2 & 18.4 & \underline{56.5} & 62.2 & -- & -- & 44.0 \\
mmE5~\cite{chen2025mme5} & 11B & \textbf{60.6} & 55.7 & 54.7 & 72.4 & -- & -- & \underline{58.6} \\
\midrule
\ours-2B & 2B & 53.6 & \underline{59.9} & 55.2 & \underline{74.6} & -- & -- & 58.2 \\
\ours-7B & 7B & \underline{56.0} & \textbf{64.7} & \textbf{58.9} & \textbf{75.7} & -- & -- & \textbf{61.6} \\
\midrule
\multicolumn{9}{c}{\textit{Fine-tuning on MMEB benchmark ($<7B$ Models)}} \\
\midrule
CLIP~\cite{radford2021learning} & 0.4B & 55.2 & 19.7 & 53.2 & 62.2 & 47.6 & 42.8 & 45.4 \\
VLM2Vec~\cite{jiang2025vlm2vec} & 4B & 54.8 & 54.9 & 62.3 & 79.5 & 66.5 & 52.0 & 60.1 \\
LLaVE~\cite{lan2025llave} & 2B & 62.1 & 60.2 & 65.2 & 84.9 & 69.4 & 59.8 & 65.2 \\
UniME~\cite{gu2025unime} & 4B & 54.8 & 55.9 & 64.5 & 81.8 & 68.2 & 52.7 & 64.2 \\
B3-2B~\cite{thirukovalluru2025b3} & 2B & \textbf{67.0} & 61.2 & \textbf{70.9} & 79.9 & 72.1 & \underline{63.1} & \underline{68.1} \\
MoCa-3B~\cite{chen2025moca} & 3B & 59.8 & \underline{62.9} & \underline{70.6} & \textbf{88.6} & \underline{72.3} & 61.5 & 67.5 \\
\midrule
\ours-2B & 2B & \underline{66.2} & \textbf{65.6} & 70.1 & \underline{85.8} & \textbf{72.9} & \textbf{65.0} & \textbf{69.5} \\ \midrule
\multicolumn{9}{c}{\textit{Fine-tuning on MMEB benchmark ($\geq7B$ Models)}} \\
\midrule
VLM2Vec~\cite{jiang2025vlm2vec} & 7B & 61.2 & 49.9 & 67.4 & 86.1 & 67.5 & 57.1 & 62.9 \\
MMRet~\cite{zhou2024megapairs} & 7B & 56.0 & 57.4 & 69.9 & 83.6 & 68.0 & 59.1 & 64.1 \\
mmE5~\cite{chen2025mme5} & 11B & 67.6 & 62.7 & 71.0 & 89.7 & 72.4 & 66.6 & 69.8 \\
LLaVE~\cite{lan2025llave} & 7B & 65.7 & 65.4 & 70.9 & \underline{91.9} & 75.0 & 64.4 & 70.3 \\
UniME~\cite{gu2025unime} & 7B & 66.8 & \underline{66.6} & 70.6 & 90.9 & 74.6 & 65.8 & 70.7 \\
B3-7B~\cite{thirukovalluru2025b3} & 7B & \textbf{70.0} & 66.5 & \underline{74.1} & 84.6 & \underline{75.9} & 67.1 & \underline{72.0} \\ 
MoCa-7B~\cite{chen2025moca} & 7B & 65.8 & 64.7 & \textbf{75.0} & \textbf{92.4} & 74.7 & \underline{67.6} & 71.5 \\
\midrule
\ours-7B & 7B & \underline{68.3} & \textbf{71.9} & 73.7 & 90.9 & \textbf{77.3} & \textbf{69.1} & \textbf{73.6} \\
\bottomrule
\end{tabular}
\vspace{-1.5em}
\end{table*}

%% file: tables/main_mbeir_small.tex
\begin{table}[t]
\centering
\caption{\textbf{Recall(\%) Results on M-BEIR.} $S$: Single modality (text or image), $M$: Multi-modality (text and image). The arrow denotes `query $\to$ target'. Full per-dataset results are in the supplementary material. \textbf{Boldface} denotes the best scores in the subset and the second-best scores are highlighted with \underline{underline}.
} 
\label{tab:main_mbeir}
\tabcolsep=0.2em
\vspace{-1em}
\resizebox{\linewidth}{!}{
\begin{tabular}{lrccccc}
\toprule
\textbf{Models} & \textbf{\# Params} &\small $S \to S$ &\small $S \to M$ &\small $M \to S$ &\small $M \to M$ & \textbf{Overall} \\
\midrule
UniIR~\cite{wei2024uniir} & 0.4B & \textbf{51.0} & \textbf{69.1} & 32.9 & 52.4 & 48.9 \\
LamRA-Ret~\cite{liu2025lamra} & 2B & 47.6 & 66.2 & \underline{41.0} & \underline{61.0} & \underline{50.0} \\
\ours-2B & 2B & \underline{49.1} & \underline{68.2} & \textbf{41.6} & \textbf{64.8} & \textbf{51.6} \\
\midrule
MM-Embed~\cite{lin2025mmembed} & 7B & 50.9 & \textbf{76.9} & 40.0 & 60.9 & 52.7 \\
LamRA-Ret~\cite{liu2025lamra} & 7B & \underline{52.9} & 71.8 & \underline{45.2} & \underline{65.0} & \underline{54.9} \\
M3Task-UEM~\cite{sharma2025multi} & 7B & \textbf{54.0} & \underline{74.9} & 41.9 & 55.7 & 53.9 \\
\ours-7B & 7B & \textbf{54.0} & 71.6 & \textbf{47.4} & \textbf{70.4} & \textbf{56.6} \\
\bottomrule
 \end{tabular}
}
\vspace{-1.5em}
\end{table}

%% file: tables/ablation_pretraining.tex
\begin{table}[t]
\footnotesize
\centering
\caption{\textbf{Pretraining dataset variants}. `None' represents fine-tuning without pretraining. We compare against other pretraining datasets (mmE5 and MegaPair) and our M3T at various scales. The 'mmE5' entry under the 'Multi-turn' section is generated using images from the single-turn mmE5 synth dataset.}
\label{tab:abl_pretraining}
\vspace{-1em}
\begin{tabular}{llrcc}
\toprule
\textbf{Pre-training} & \textbf{Dataset} & \textbf{Samples} & \begin{tabular}[c]{@{}c@{}} \textbf{ZS} \\ \textbf{MMEB}\end{tabular} & \begin{tabular}[c]{@{}c@{}} \textbf{FT} \\ \textbf{MMEB}\end{tabular} \\ \midrule
\multirow{3}{*}[0.1em]{Single-turn} & None & & -- & 68.5 \\
& mmE5~\cite{chen2025mme5} & 0.6M & 55.6 & 68.6 \\
& MegaPairs~\cite{zhou2024megapairs} & 26M & 41.5 & 68.7 \\
\midrule
\multirow{4}{*}[0.1em]{\begin{tabular}[c]{@{}c@{}} Multi-turn \\ \ours-\S\ref{sec:muco_pretraining} \end{tabular}} & mmE5~\cite{chen2025mme5} & 0.6M & 57.0 & 69.0 \\
& M3T (20\%) & 1M & 57.1 & 69.0 \\
& M3T (60\%) & 3M & 57.7 & 69.2 \\
& M3T & 5M & 58.2 & 69.5 \\ \bottomrule
\end{tabular}
\vspace{-2em}
\end{table}

%% file: tables/ablation_each_task.tex
\begin{table}[t]
\footnotesize
\centering
\caption{\textbf{Ablation study for data composition.} `All' means the full dataset. Subsequent rows show performance after sequentially removing task categories: Classification ($-$ CLS), followed by global VQA ($-$ global VQA; holistic understanding), local VQA ($-$ local VQA; localized details), and creative VQA ($-$ creative VQA; creative reasoning). All results are in the zero-shot setting on MMEB. GRD stands for visual grounding. The final row is the result of training only on the remaining Retrieval (RET) data.}
\vspace{-1em}
\label{tab:abl_each_task}
\begin{tabular}{lccccc}
\toprule
\textbf{Setup} & \textbf{CLS} & \textbf{VQA} & \textbf{RET} & \textbf{GRD} & \textbf{Overall} \\ \midrule
All & 53.6 & 59.9 & 55.2 & 74.6 & 58.2 \\ \midrule
$-$ CLS & 51.7 & 59.2 & 54.6 & 72.5 & 56.9 \\
$-$ global VQA & 51.1 & 57.6 & 54.6 & 71.5 & 56.3 \\
$-$ local VQA & 50.8 & 56.1 & 54.1 & 70.2 & 55.5 \\
$-$ creative VQA & 50.2 & 55.7 & 53.7 & 69.8 & 55.1 \\ \bottomrule
\end{tabular}
\vspace{-1em}
\end{table}

%% file: tables/ablation_multiturn_batchsize.tex
\begin{table}[t]
\footnotesize
\centering
\caption{\textbf{Analysis of scaling batch size vs. turns.} We compare the baseline method (\#turns=1) with increasing batch sizes (1024 to 8192) against our proposed \ours\ method (\#turns $\ge 2$) with an increasing number of turns at a fixed batch size 1024 in a zero-shot setting on MMEB. For the 2-turn and 4-turn settings, pairs were randomly sampled from the 7 available pairs. The `effective batch' refers to the total number of query-target pairs (\#batch $\times$ \#turns), which amplifies the learning signal by providing more positive and negative pairs for the contrastive loss.
}
\label{tab:abl_multiturn_batchsize}
\vspace{-1em}
\begin{tabular}{lccrrr}
\toprule
\textbf{\#turns} & \textbf{\#batch} & \textbf{\#effective batch} & \textbf{PFLOPs} & \textbf{MMEB} \\ \midrule
1 & 1024 & 1024 & 17.5 & 57.1 \\
1 & 2048 & 2048 & 35.1 & 57.3 \\
1 & 4096 & 4096 & 70.2 & 57.4 \\
1 & 7168 & 7168 & 122.7 & 57.5 \\
1 & 8192 & 8192 & 140.4 & 57.8 \\ \midrule
2 & 1024 & 2048 & 17.6 & 57.4 \\
4 & 1024 & 4096 & 17.7 & 57.7 \\
7 & 1024 & 7168 & 18.0 & 58.2 \\ \bottomrule
\end{tabular}
\vspace{-2em}
\end{table}

%% file: tables/ablation_causal_attention.tex
\begin{table}[t]
\footnotesize
\centering
\caption{\textbf{Impact of compounded supervision.} Preventing causal attention to attend earlier contexts leads to a performance drop, supporting the importance of compounded supervision. }
\label{tab:abl_causal_attention}
\vspace{-1em}
\begin{tabular}{ccc}
\toprule
\textbf{Setup} & \textbf{ZS MMEB} & \textbf{FT MMEB} \\ \midrule
w/o compounded supervision & 57.3 & 68.4 \\
w/ compounded supervision & 58.2 & 69.5 \\ \bottomrule
\end{tabular}
\vspace{-1em}
\end{table}

%% file: tables/ablation_intra_mask.tex
\begin{table}[t]
\footnotesize
\centering
\caption{\textbf{Impact of logit masking.} While beneficial in pre-training due to its diverse multi-turn data, this strategy is critical for fine-tuning, as it prevents a severe performance collapse caused by incorrectly treating self-augmented pairs as negatives}
\label{tab:abl_intra_mask}
\vspace{-1em}
\begin{tabular}{cccc}
\toprule
\multicolumn{2}{c}{\textbf{Logit masking}} &  & \\
\textbf{Pretraining} & \textbf{Fine-tuning}  & \textbf{ZS MMEB} & \textbf{FT MMEB} \\ \midrule
- & - & 57.7 & 31.1 \\
- & \checkmark & 57.7 & 69.2 \\
\checkmark & - & 58.2 & 30.9 \\
\checkmark & \checkmark & 58.2 & 69.5 \\ \bottomrule
\end{tabular}
\vspace{-1.5em}
\end{table}

%% file: tables/ablation_finetuning_design.tex
\begin{table}[t]
\footnotesize
\centering
\caption{\textbf{Ablation study of the subsequent turn design for fine-tuning on single-pair dataset.} We analyze the counterpart masking ratio, the importance of reconstruction guidance (compared to no guidance or simple rephrasing), and find that the choice of image captioning model has a minimal impact}
\label{tab:abl_finetuning_design}
\vspace{-1em}
\begin{tabular}{lc}
\toprule
\textbf{Setup} & \textbf{MMEB} \\ \midrule
Counterpart masking ratio 25\% &  69.3 \\
Counterpart masking ratio 50\% & 69.5 \\
Counterpart masking ratio 75\% &  68.9 \\ \midrule
Rephrasing template & 68.5 \\
Without reconstruction guidance & 69.0 \\
Image captioning (BLIP Large~\cite{li2022blip}) & 69.4 \\
Image captioning (Qwen2-VL-7B~\cite{wang2024qwen2}) & 69.5 \\ \bottomrule
\end{tabular}
\vspace{-1.5em}
\end{table}

%% file: appendix_preprint.tex
\appendix




%


\def\paperID{****} 
\def\confName{CVPR}
\def\confYear{2026}


\setcounter{table}{0}
\renewcommand{\thetable}{\Alph{table}}
\setcounter{figure}{0}
\renewcommand{\thefigure}{\Alph{figure}}
\setcounter{section}{0}
\renewcommand\thesection{\Alph{section}}

\maketitlesupplementary

\section*{Appendix}
We include additional materials in this document.
\begin{itemize}
    \item \hyperref[apdx_a]{\textbf{Section A:} Clarification after the initial submission}
    \item \hyperref[apdx_b]{\textbf{Section B:} Detailed comparison of pretraining datasets}
    \item \hyperref[apdx_c]{\textbf{Section C:} Details of benchmark datasets for fine-tuning}
    \item \hyperref[apdx_d]{\textbf{Section D:} Details of synthesizing our M3T dataset}
    \item \hyperref[apdx_e]{\textbf{Section E:} Additional ablation studies}
    \item \hyperref[apdx_f]{\textbf{Section F:} Extended main tables}
    \item \hyperref[apdx_g]{\textbf{Section G:} M3T examples}
\end{itemize}

\section{Additional clarification for Tab.~5}
\label{apdx_a}
Regarding the batch size scaling experiments in Tab.~5 of the main paper, we clarify that the learning rates were scaled according to the batch size: $5e^{-5}$ for batch size 2,048, $1e^{-4}$ for 4,096, and $2e^{-4}$ for batch sizes 7,168 and 8,192.

\section{Detailed comparison of pretraining datasets}
\label{apdx_b}
\input{tables/appendix_details_pretraining_datasets}
We include this section to further substantiate the effectiveness and efficiency of our M3T dataset, providing a more detailed analysis than the main paper. As shown in \cref{tab:apdx_details_pretraining_datasets}, we present a comprehensive statistical and performance comparison with mmE5 and MegaPairs. This comparison highlights M3T's superior efficiency (evident in its simplified IT-2-T structure and zero hard negatives) and its effectiveness (achieving state-of-the-art zeroshot and fine-tuning performance on the MMEB and M-BEIR benchmarks).

\noindent\textbf{Differences in meta-Task composition.}
The datasets differ in their meta-task composition. Both mmE5 and our M3T include samples for CLS, VQA, and RET, which are core categories from the MMEB benchmark. Notably, M3T features a highly diverse VQA set (25.5M pairs) categorized into global, local, and creative sub-tasks. In contrast, MegaPairs consists solely of retrieval samples, specifically for the IT-2-I modality.
This compositional difference is clearly reflected in the zeroshot (ZS) MMEB benchmark performance. As shown in \cref{tab:apdx_details_pretraining_datasets}, M3T and mmE5, which align with the MMEB benchmark's core categories, significantly outperform MegaPairs in the ZS setting (Overall scores: 58.2 and 55.6 vs. 41.5). Interestingly, this trend shifts after fine-tuning (FT), where MegaPairs (68.7) surpasses mmE5 (68.6), suggesting that Overall FT performance correlates strongly with data scale.
Therefore, we scaled M3T to 5M images, achieving the highest performance in both zeroshot and fine-tuning. One notable observation is that despite M3T's large volume of VQA data, it does not lead to disproportionately high VQA performance at the expense of other tasks. Instead, it contributes to a robust improvement across all categories. We attribute this balanced enhancement to our \ours\ framework, which effectively leverages this task diversity to refine the quality of the initial turn's embedding.


\noindent\textbf{Differences in modality composition and robust generalization.} A unique characteristic of M3T is its exclusive reliance on the IT-2-T (Image+Text query to Text target) modality structure, unlike mmE5 or MegaPairs which utilize diverse combinations. Despite this constraint, M3T demonstrates remarkable generalization in the M-BEIR benchmark. In the zeroshot setting, M3T achieves the highest Overall score (37.8), surpassing both mmE5 (37.3) and MegaPairs (35.7). Specifically, M3T secures the best performance in $S \to S$ (38.2) and notably in $M \to M$ tasks (48.3). While M3T shows a slightly lower score in the zeroshot $M \to S$ setting (27.0) compared to MegaPairs (29.6), this gap is effectively bridged after fine-tuning, where M3T achieves comparable performance ($45.5$ vs. $45.8$) and dominates across all other metrics.
Crucially, the robustness of our learned representations is highlighted in the global retrieval setting, where the candidate pool includes all datasets combined. As shown in Table~\ref{tab:apdx_details_pretraining_datasets}, while the inclusion of massive distractors causes a performance drop across all models, training with M3T consistently outperforms baselines in the global setting (Overall 17.4 vs. 15.8 for mmE5). This suggests that training with M3T learns a globally discriminative embedding space. Even without domain-specific tuning, our model effectively mitigates inter-task interference and maintains semantic separability against distractors from heterogeneous tasks, a capability that is further amplified after fine-tuning (Global Overall 51.6).

\noindent\textbf{Computational Efficiency and Training Statistics.} 
To analyze efficiency from a computational perspective, we list the statistical characteristics observed during actual training in \cref{tab:apdx_details_pretraining_datasets}. All metrics were measured using the Qwen2-VL-2B model with a batch size of 1,024, specifically breaking down the computational load into query, positive target, and explicit hard-negative batches. A key distinction is that M3T does not utilize explicitly mined hard negatives. Consequently, M3T exhibits a lower total token count compared to mmE5, leading to a reduced time per iteration (8.41s for M3T vs. 15.33s for mmE5). It is crucial to note that the token counts reported for M3T encompass the simultaneous processing of all seven multi-turn pairs, whereas the metrics for other datasets only reflect the processing of single pairs. MegaPairs shows the lowest token counts, GFLOPs, and time per iteration; this is primarily because it employs a fixed image resolution of $512 \times 512$, whereas both M3T and mmE5 support variable resolutions. Regarding total training time for one epoch, MegaPairs takes the longest (53.4 hours) due to its massive scale, followed by M3T (11.6 hours), with mmE5 being the fastest (2.3 hours) due to its smaller dataset size. However, when normalizing for scale, the efficiency of M3T becomes evident. If mmE5 were scaled to match the volume of MegaPairs or M3T, its estimated training time would surge from 2.3 hours to approximately 109 hours and 21.2 hours (nearly double the time required for M3T), respectively. This confirms that the M3T framework is explicitly designed to maximize both performance and computational efficiency.

\section{Details of benchmark datasets for fine-tuning.}
\label{apdx_c}
\input{tables/appendix_benchmark_datasets}

\begin{figure*}[t]
\centering
\includegraphics[width=\linewidth]{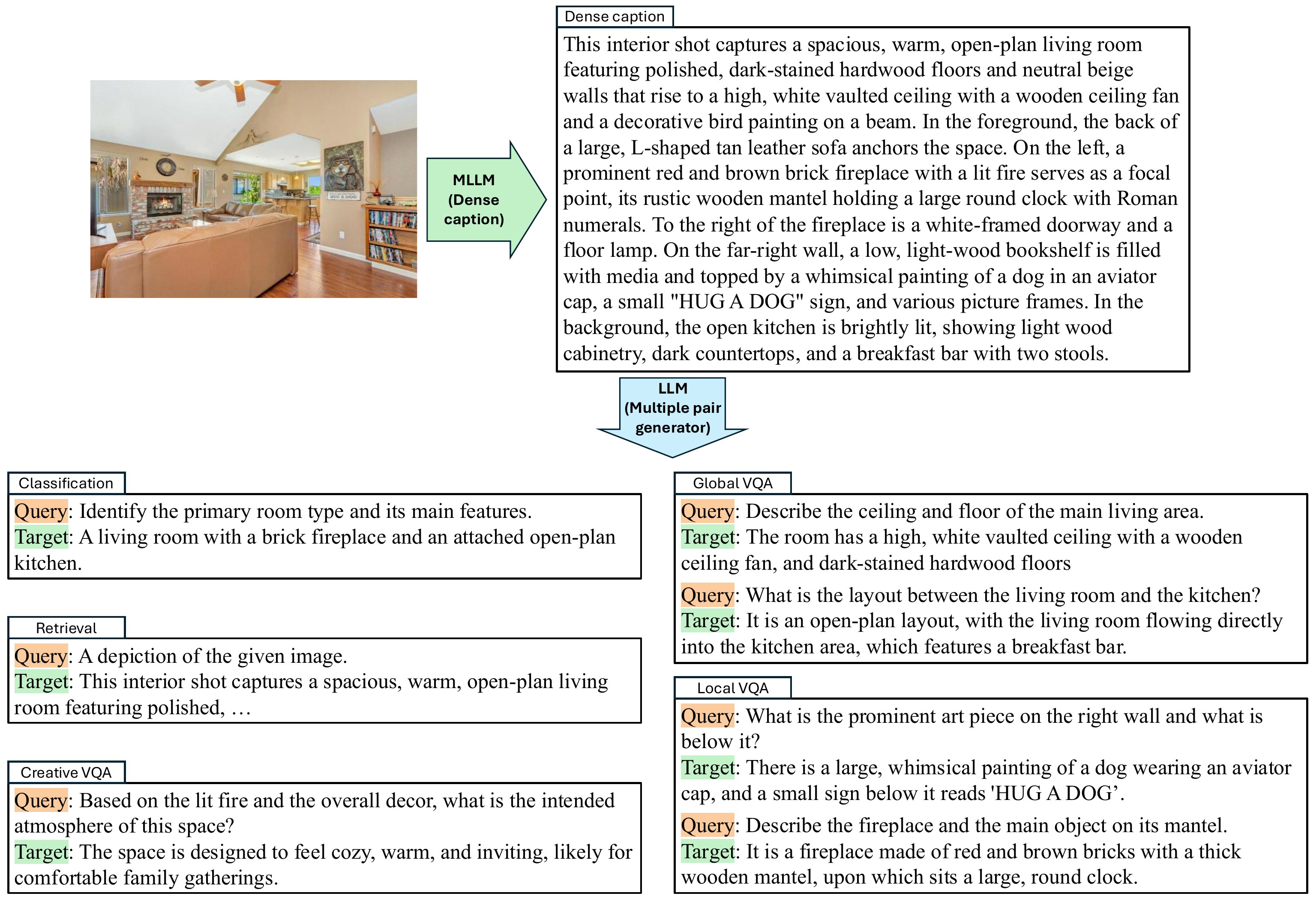}
\caption{\textbf{Overview of the M3T Dataset Synthesis Pipeline.} 
The data synthesis proceeds in two stages. First, an MLLM (Qwen2.5-VL-75B) is used to generate a comprehensive, objective dense caption of an image's rich visual information. Second, an LLM (OSS-20B) takes the dense caption as input to synthesize a diverse set of text-only query-positive pairs aligned with seven meta-tasks. Specifically, these include one for classification, one retrieval, and five distinct VQA pairs (two global VQA, two local VQA, and one creative VQA). For the retrieval task, the dense caption itself serves as the positive target, which is abbreviated with ``..." for clarity in the figure.
}
\label{fig:data_synthesis}
\vspace{-1.5em}
\end{figure*}

\begin{figure}[t]
\centering
\includegraphics[width=\linewidth]{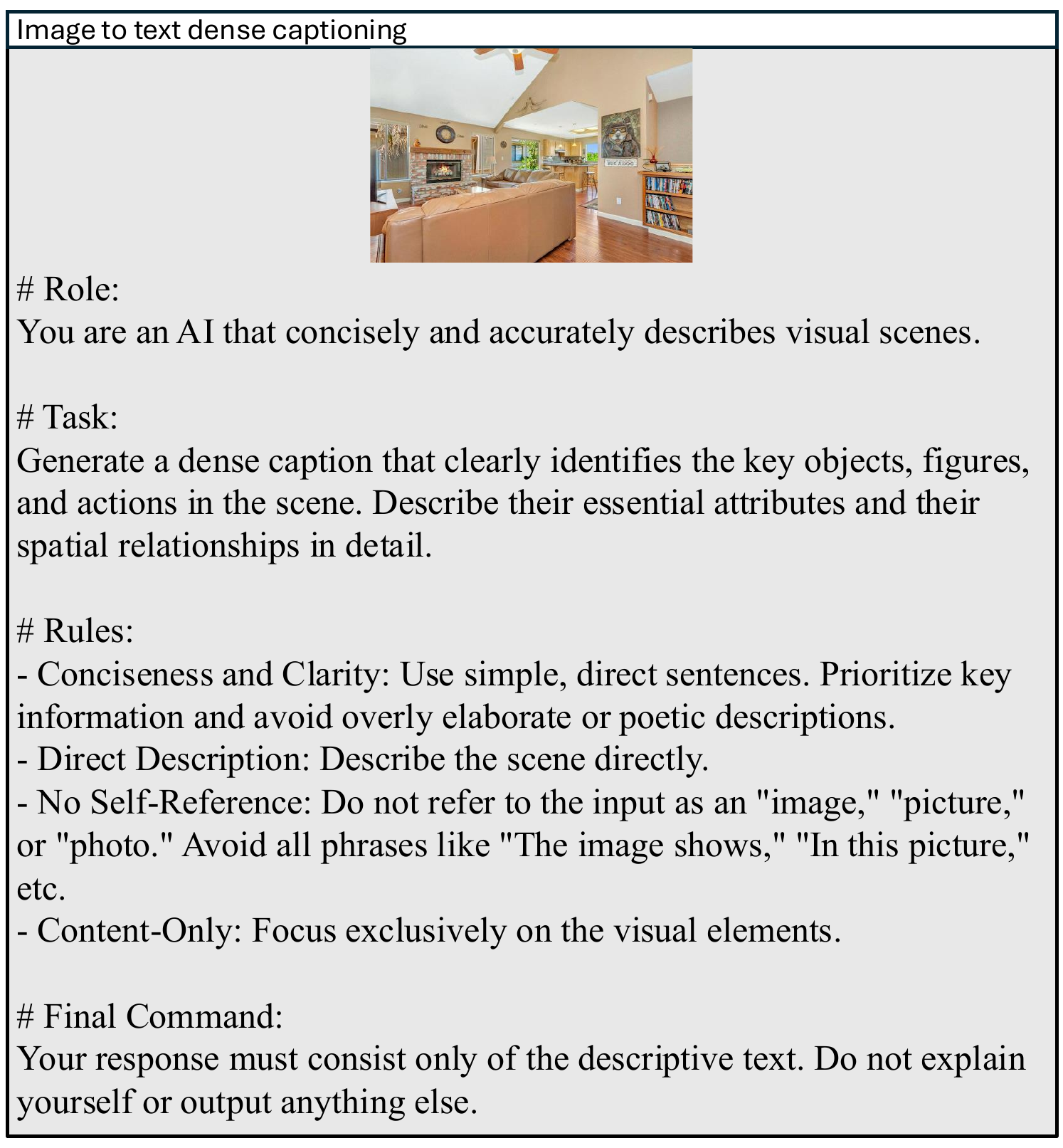}
\caption{\textbf{Prompt used for dense image captioning with an MLLM.} The prompt is structured into four sections: Role, Task, Rules, and Final command. The core objective is to instruct the MLLM to generate a dense, objective description capturing all key visual information.
}
\label{fig:prompt_i2t}
\vspace{-1.5em}
\end{figure}

In this work, we evaluate our proposed method and compare it with previous methods using two representative benchmarks for universal multimodal embeddings: MMEB and M-BEIR.
As detailed in \cref{tab:MMEB_datasets}, MMEB comprises four meta-tasks: Classification, Visual Question Answering, Retrieval, and Visual Grounding, consisting of 10, 10, 12, and 4 datasets, respectively.
As shown in \cref{tab:MBEIR_datasets}, M-BEIR consists of 16 datasets. Unlike MMEB, M-BEIR focuses exclusively on retrieval tasks, serving as a specialized benchmark to analyze retrieval performance across diverse modalities.

\section{Details of synthesizing our M3T dataset}
\label{apdx_d}
We introduce the \textbf{M3T} (Multi-modal multi-turn) dataset, a large-scale dataset designed for pretraining robust multimodal embedding models. Here, we describe the synthesis process of our M3T dataset.




The M3T data synthesis pipeline is intentionally straightforward as depicted in \cref{fig:data_synthesis}. The construction process proceeds in two main steps. The first step involves dense image captioning using a state-of-the-art MLLM, Qwen2.5-VL-75B~\cite{bai2025qwen2.5}.
For the first step, we randomly sample 5 million images from DataComp~\cite{gadre2023datacomp}, selecting only images where at least one spatial dimension is 512 pixels or larger, a threshold established empirically.
We observe this is a crucial factor for the MLLM to accurately recognize small-sized objects and perform Optical Character Recognition (OCR) on text within the images during the first step.
Guided by a prompt (\cref{fig:prompt_i2t}), the MLLM is instructed to produce objective, direct descriptions that identify key objects, their essential attributes, and their spatial relationships. This initial step effectively distills the rich visual information of an image into a dense caption, focusing on observable content.

\begin{figure*}[t]
\centering
\includegraphics[width=\linewidth]{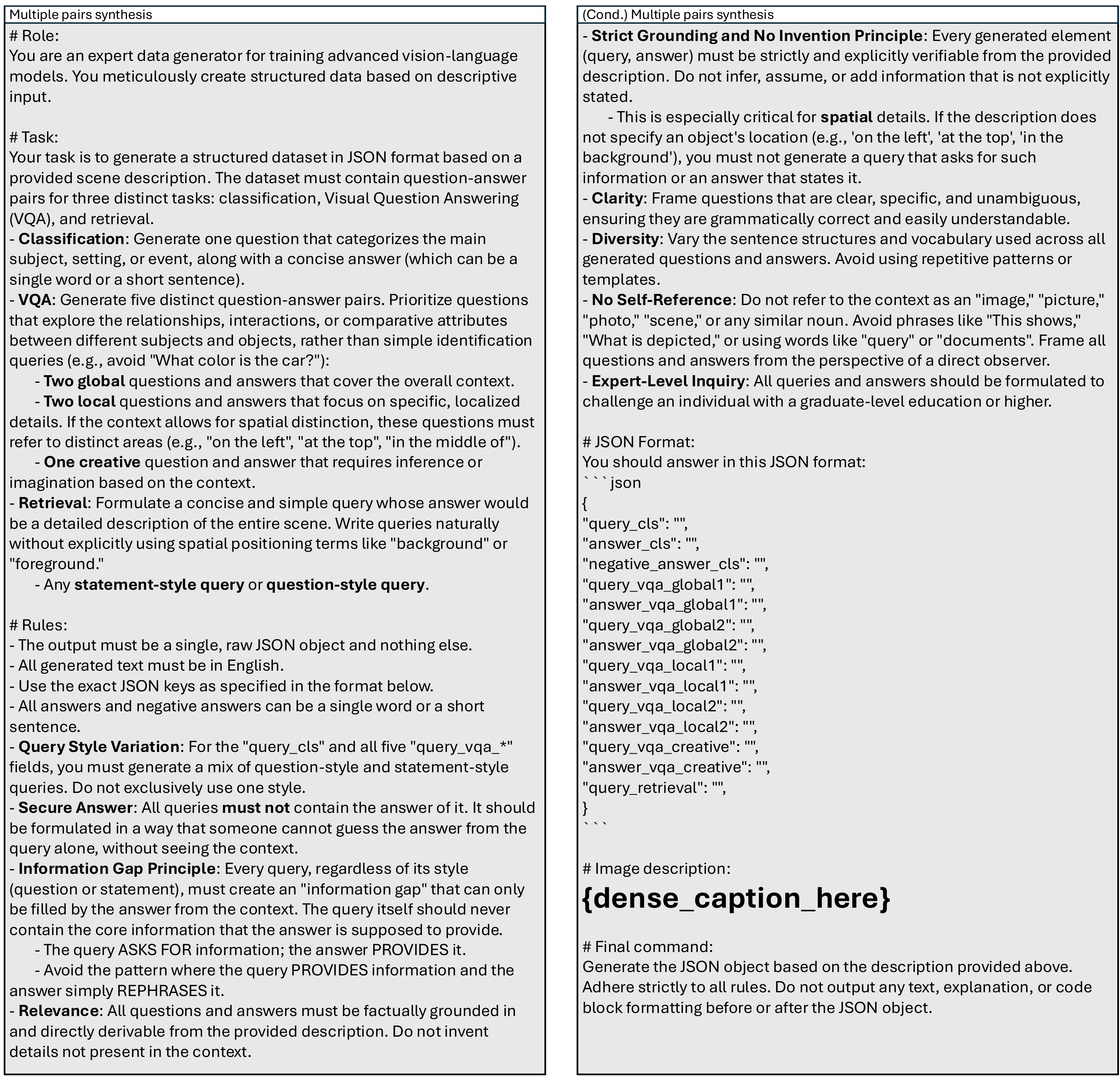}
\caption{\textbf{Prompt used for synthesizing multiple query-target pairs with the LLM.} For clarity, the prompt is plotted in two columns. Unlike the prompt for dense image captioning (\cref{fig:prompt_i2t}), this prompt is empirically longer and applies more detailed, structured rules to generate the seven distinct pairs from an input dense caption.
}
\label{fig:prompt_t2d}
\vspace{-1.5em}
\end{figure*}

In the second step, we use the generated dense captions as input to a LLMs, OSS-20B~\cite{agarwal2025gpt}, to synthesize a diverse set of query-positive pairs for each image. We design these synthesized pairs to align with the core categories of the MMEB benchmark. Specifically, we generate the following pairs for each image: 1) a query for classifying the image's dominant semantic content, 2) one retrieval query, for which the dense caption serves as the positive target, and 3) five distinct Visual Question Answering (VQA) pairs. The VQA pairs were designed to encourage different reasoning abilities: two pairs require a holistic understanding of the entire scene, two pairs require focusing on localized details, and one creative pair demands reasoning beyond literal description. Notably, this entire set of query-positive pairs is synthesized in a single pass using one comprehensive prompt (\cref{fig:prompt_t2d}) for the LLM.

This two-stage approach is intentional, as it is designed to create an efficient and extensible pipeline. By isolating the computationally intensive image-to-text conversion as a distinct, one-time preliminary step, the subsequent synthesis of query-positive pairs becomes a flexible and inexpensive text-only operation. Consequently, our 5-million-image dataset yields a total of 35 million query-positive pairs with 5 million images.

\input{tables/appendix_additional_finetuning_design}
\input{tables/ablation_finetuning_combination}
\input{tables/appendix_only_finetuning}

\section{Additional ablation studies}
\label{apdx_e}
\noindent\textbf{Impact of pretraining and fine-tuning stages.} To investigate the individual contributions of the pretraining and fine-tuning stages, we conduct an ablation study reported in \cref{tab:apdx_only_finetuning_mmeb} for MMEB and \cref{tab:apdx_only_finetuning_mbeir} for M-BEIR. Interestingly, our results demonstrate that the \ours\ fine-tuning strategy alone is highly effective; even without pretraining, it surpasses previous state-of-the-art methods. For instance, on MMEB, \ours-2B (fine-tuning only) achieves 68.4\%, outperforming the fully trained B3-2B (68.1\%). Similarly, \ours-7B (fine-tuning only) reaches 72.6\%, exceeding B3-7B (72.0\%). A similar trend is observed on M-BEIR, where our fine-tuning-only models consistently outperform strong baselines like LamRA-Ret. Furthermore, incorporating the pretraining stage with our M3T dataset provides a significant performance boost. This confirms that the multi-turn learning signals are beneficial in both stages. Consequently, the combined effect of leveraging the rich, dialogue-driven context from M3T during pretraining and the adaptive multi-turn reconstruction during fine-tuning maximizes the model's representational power, leading to the most robust performance.

\noindent\textbf{Additional ablation study of the subsequent turn design for fine-tuning on single-pair dataset.} We present an additional ablation study on subsequent turn and token design in \cref{tab:apdx_additional_finetuning_design}. The `self-reconstruction template', which utilizes the initial turn for its subsequent turn, achieves a score of 68.5, matching the `rephrasing template' result in Tab.~8 from the main paper. This suggests that self-reconstruction functions as a semantic refinement process similar to rephrasing. Furthermore, substituting the special mask token with simple text patterns yields comparable performance. This demonstrates that the model can effectively interpret the reconstruction instruction from the prompt context alone during fine-tuning.

\noindent\textbf{Various augmented data combination for fine-tuning.} We analyze the impact of different pair combinations in Eq. 6. \cref{tab:abl_finetuning_pair_combinations} shows the result. The baseline (Case 1), which uses only the initial $(q, p)$ pairs, achieves 68.1. Interestingly, using only one of the subsequent pairs (Cases 2-4) results in slightly worse performance (67.7-68.0) than the baseline. This suggests that while the subsequent turn provides accumulative supervision, this signal alone is insufficient to effectively train the initial embedding (which is used for inference). However, performance significantly improves (e.g., 69.1 in Case 5) when the initial $(q, p)$ pair is combined with any augmented pair (Cases 5-7). This validates our hypothesis: while the initial embedding is explicitly trained, the accumulative supervision from the subsequent augmented turns acts as a powerful refiner.
As expected, Cases 8-10 perform worse than Cases 5-7 because they omit the crucial training signal for the initial turn.
As expected, our full approach (Case 11), which utilizes all four combinations, achieves the highest performance (69.5).

\section{Extended main tables}
\label{apdx_f}
In this section, we present the comprehensive experimental results for the MMEB and M-BEIR benchmarks, expanding upon the summarized findings in the main text. Due to space constraints, the main paper reported only aggregated performance metrics. Here, we provide the full breakdown: \cref{tab:apdx_extended_mmeb} details the Precision@1 scores for all 36 individual datasets within the MMEB benchmark, covering both In-Distribution and Out-of-Distribution splits. \cref{tab:apdx_extended_mbeir} presents the detailed retrieval performance for the M-BEIR benchmark, reporting Recall metrics (Recall@5 or Recall@10) for each of the 16 datasets across diverse retrieval tasks. We also report quantitative results (M-BEIR$_\text{local}$) evaluated using local candidate pools specific to each of the 16 datasets in \cref{tab:apdx_extended_mbeir_local}.

\section{M3T examples}
\label{apdx_g}
\begin{figure*}[t]
\centering
\includegraphics[width=\linewidth]{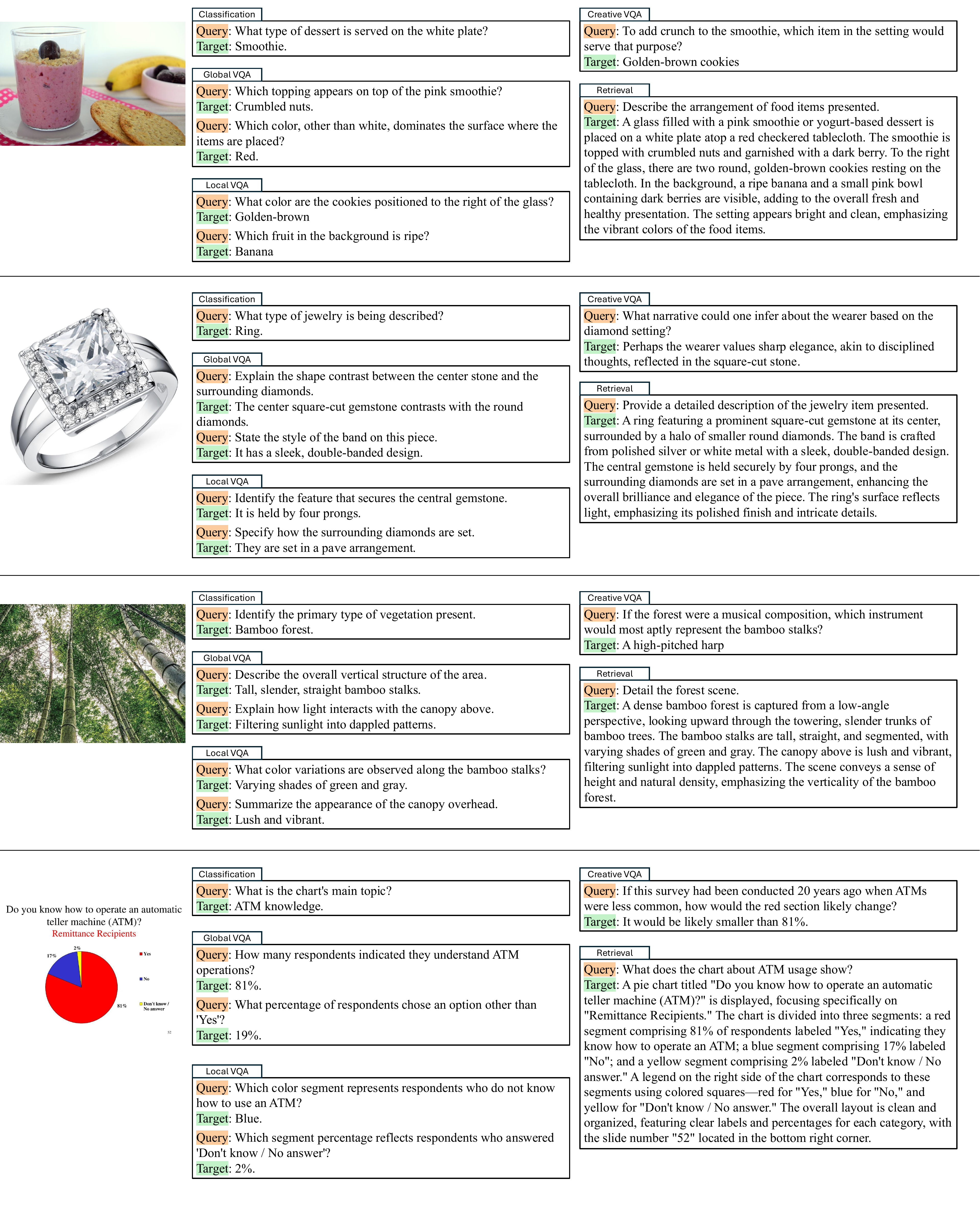}
\caption{\textbf{Examples of our M3T.}
}
\label{fig:M3T_examples1}
\vspace{-1.5em}
\end{figure*}

\begin{figure*}[t]
\centering
\includegraphics[width=\linewidth]{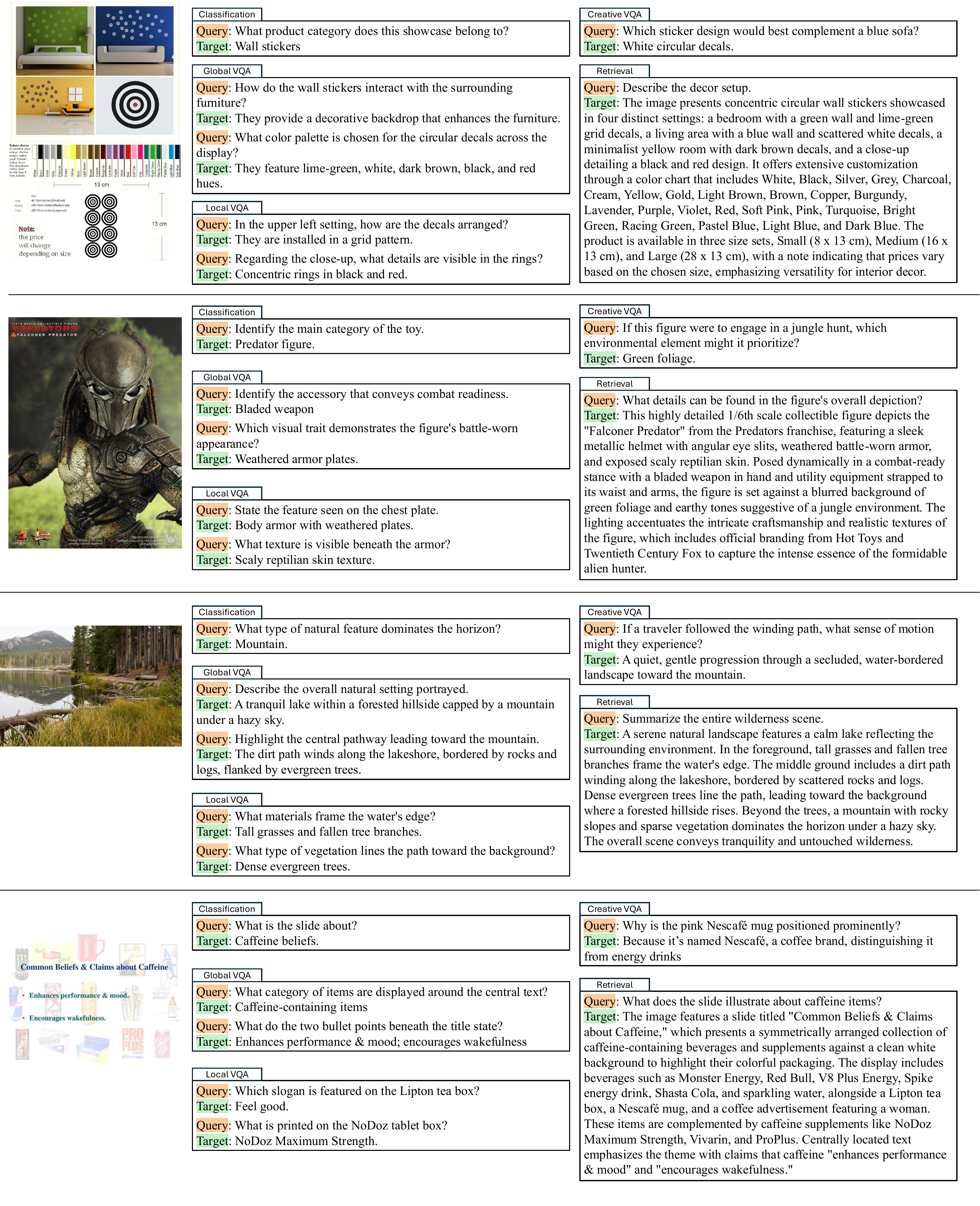}
\caption{\textbf{Examples of our M3T.}
}
\label{fig:M3T_examples2}
\vspace{-1.5em}
\end{figure*}

We present examples of our dataset in \cref{fig:M3T_examples1} and \cref{fig:M3T_examples2}.

\input{tables/appendix_extended_mbeir}
\input{tables/appendix_extended_mmeb}

\clearpage
\clearpage


%% file: tables/appendix_details_pretraining_datasets.tex
\begin{table}[t]
\centering
\caption{\textbf{Detailed comparison of pretraining datasets.} We compare our M3T with mmE5~\cite{chen2025mme5} and MegaPairs~\cite{zhou2024megapairs} across dataset statistics, training efficiency, and downstream performance. Metrics marked with $\dagger$ are measured using the \ours-2B model with 1024 batch size on 32 A100 GPUs. For M-BEIR, values in parentheses indicate results evaluated using local candidate pools specific to each of the 16 datasets.}
\vspace{-1em}
\label{tab:apdx_details_pretraining_datasets}
\resizebox{0.92\linewidth}{!}{

\begin{tabular}{lccc}
\toprule
\textbf{Metric} & \textbf{mmE5} & \textbf{MegaPairs} & \textbf{M3T (ours)} \\ 
\midrule
\multicolumn{4}{c}{\textbf{\# of Samples}} \\
\midrule
\# of samples & 560,000 & 26,235,105 & 5,103,183 \\
\midrule
\multicolumn{4}{c}{\textbf{\# of Pairs per Meta-task}} \\
\midrule
\# of \textbf{All} pairs & 560,000 & 26,235,105 & 5,103,183 \\
\# of \textbf{CLS} pairs & 140,000 & 0 & 5,103,183 \\
\# of \textbf{VQA} pairs & 140,000 & 0 & 25,515,915 \\
\# of \textbf{RET} pairs & 280,000 & 26,235,105 & 5,103,183 \\
\midrule
\multicolumn{4}{c}{\textbf{\# of Pairs per modality}} \\
\midrule
\# of T-2-T & 0 & 0 & 0\\
\# of T-2-I & 14,090 & 0 & 0\\
\# of T-2-IT & 14,081 & 0 & 0\\
\# of I-2-T & 224,217 & 0 & 0\\
\# of I-2-I & 27,988 & 0 & 0\\
\# of IT-2-T & 195,783 & 0 & 35,722,281\\
\# of IT-2-I & 56,185 & 26,235,105 & 0\\
\# of IT-2-IT & 27,656 & 0 & 0\\
\midrule
\multicolumn{4}{c}{\textbf{Batch metrics and training time for 1024 batch}} \\
\midrule
Avg batch Qry tokens$\dagger$ & 1,318,912 & 447,488 & 1,252,352 \\
Avg batch Pos tokens$\dagger$ & 918,528 & 432,128 & 596,992 \\
Avg batch HN tokens $\dagger$ & 828,416 & 361,472 & 0 \\
GFLOPs per Iteration$\dagger$ & 37.5 & 13.7 & 18.6 \\
Second per Iteration$\dagger$ & 15.33 & 7.51 & 8.41 \\
Total Iterations$\dagger$ & 547 & 25,620 & 4,984 \\
Total Hours$\dagger$ & 2.3 & 53.4 & 11.6 \\
\midrule
\multicolumn{4}{c}{\textbf{Zeroshot setting on \textit{MMEB} benchmark}} \\
\midrule
ZS MMEB (CLS) & 51.1 & 50.4 & 53.6 \\
ZS MMEB (VQA) & 58.8 & 21.6 & 59.9 \\
ZS MMEB (RET) & 53.4 & 45.2 & 55.2 \\
ZS MMEB (GRD) & 65.2 & 57.5 & 74.6 \\
ZS MMEB (Overall) & 55.6 & 41.5 & 58.2 \\
\midrule
\multicolumn{4}{c}{\textbf{Fine-tuning on \textit{MMEB} benchmark}} \\
\midrule
FT MMEB (CLS) & 65.4 & 65.9 & 66.2 \\
FT MMEB (VQA) & 65.6 & 64.2 & 65.6 \\
FT MMEB (RET) & 69.5 & 69.7 & 70.1 \\
FT MMEB (GRD) & 81.1 & 83.7 & 85.8 \\
FT MMEB (Overall) & 68.6 & 68.7 & 69.5 \\
\midrule
\multicolumn{4}{c}{\textbf{Zeroshot setting on \textit{M-BEIR} benchmark}} \\
\midrule
ZS M-BEIR (\small $S \to S$) & 12.7(36.8) & 11.7(34.6) & 15.1(38.2) \\
ZS M-BEIR (\small $S \to M$) & 21.7(48.6) & 14.6(45.9) & 21.7(47.5) \\
ZS M-BEIR (\small $M \to S$) & 9.0(27.8) & 16.7(29.6) & 11.0(27.0) \\
ZS M-BEIR (\small $M \to M$) & 35.9(47.0) & 32.6(42.1) & 35.4(48.3) \\
ZS M-BEIR (Overall) & 15.8(37.3) & 15.9(35.7) & 17.4(37.8) \\
\midrule
\multicolumn{4}{c}{\textbf{Fine-tuning on \textit{M-BEIR} benchmark}} \\
\midrule
FT M-BEIR (\small $S \to S$) & 47.1(50.6) & 48.1(50.6) & 49.1(51.4) \\
FT M-BEIR (\small $S \to M$) & 65.8(66.5) & 66.0(66.9) & 68.2(68.3) \\
FT M-BEIR (\small $M \to S$) & 40.9(45.5) & 42.2(45.8) & 41.6(45.5) \\
FT M-BEIR (\small $M \to M$) & 61.5(65.8) & 62.8(66.8) & 64.8(68.7) \\
FT M-BEIR (Overall) & 49.7(53.2) & 49.7(53.5) & 51.6(54.2) \\
\bottomrule
\end{tabular}
}

\vspace{-1em}
\end{table}

%% file: tables/appendix_benchmark_datasets.tex
\begin{table}[t]
\centering
\caption{\textbf{MMEB datasets.} \textbf{Boldface} indicates out-of-distribution evaluation data, while the others represent in-distribution evaluation.}
\label{tab:MMEB_datasets}
\footnotesize
\begin{tabularx}{\linewidth}{l X}
    \toprule
    \textbf{Tasks} & \textbf{Datasets} \\
    \midrule
    Classification & ImageNet-1K~\cite{deng2009imagenet},   N24News~\cite{wang2022n24news}, HatefulMemes~\cite{kiela2020hatefulmemes}, VOC2007~\cite{everingham2015voc2007}, SUN397~\cite{xiao2010sun397}, \textbf{ImageNet-A}~\cite{hendrycks2021imageneta}, \textbf{ImageNet-R}~\cite{hendrycks2021imagenetr}, \textbf{Place365}~\cite{zhou2017places365},  \textbf{ObjectNet}~\cite{barbu2019objectnet}, \textbf{Country-211}~\cite{radford2021learning} \\ \midrule
    VQA & OK-VQA~\cite{marino2019okvqa}, A-OKVQA~\cite{schwenk2022aokvqa}, DocVQA~\cite{mathew2021docvqa}, InfographicsVQA~\cite{mathew2022infographicvqa}, ChartQA~\cite{masry2022chartqa}, Visual7W-telling~\cite{zhu2016visual7w}, \textbf{ScienceQA}~\cite{lu2022scienceqa}, \textbf{VizWiz}~\cite{gurari2018vizwiz}, \textbf{TextVQA}~\cite{singh2019textvqa}, \textbf{GQA}~\cite{hudson2019gqa} \\ \midrule
    Retrieval & VisDial~\cite{das2017visdial}, CIRR~\cite{liu2021cirr}, VisualNews~\cite{liu2021visualnews}, MSCOCO~\cite{lin2014mscoco}, NIGHTS~\cite{fu2023nights}, WebQA~\cite{chang2022webqa}, \textbf{FashionIQ}~\cite{wu2021fashioniq}, \textbf{Wiki-SS-NQ}~\cite{ma2024wikissnq}, \textbf{OVEN}~\cite{hu2023oven}, \textbf{EDIS}~\cite{liu2023edis} \\ \midrule
    Visual Grounding & MSCOCO~\cite{lin2014mscoco}, \textbf{RefCOCO}~\cite{kazemzadeh2014refcoco}, \textbf{RefCOCO-matching}~\cite{kazemzadeh2014refcoco}, \textbf{Visual7W-pointing}~\cite{zhu2016visual7w} \\
    \bottomrule
\end{tabularx}
\end{table}

\begin{table}[t]
\centering
\caption{\textbf{M-BEIR datasets.} $S$: Single modality (text or image), $M$: Multi-modality (text and image).}
\label{tab:MBEIR_datasets}
\footnotesize
\begin{tabularx}{\linewidth}{l X}
    \toprule
    \textbf{Tasks} & \textbf{Datasets} \\
    \midrule
    \small$S \to S$ & VisualNews(\textit{T2I})~\cite{liu2021visualnews}, MSCOCO(\textit{T2I})~\cite{lin2014mscoco}, Fashion-200K(\textit{T2I})~\cite{han2017fashion200k}, WebQA(\textit{T2T})~\cite{chang2022webqa}, VisualNews(\textit{I2T})~\cite{liu2021visualnews}, MSCOCO(\textit{I2T})~\cite{lin2014mscoco}, Fashion-200K(\textit{I2T})~\cite{han2017fashion200k}, NIGHTS(\textit{I2I})~\cite{fu2023nights} \\ \midrule
    \small$S \to M$ & EDIS(\textit{T2IT})~\cite{liu2023edis}, WebQA(\textit{T2IT})~\cite{chang2022webqa} \\ \midrule
    \small$M \to S$ & OVEN(\textit{IT2T})~\cite{hu2023oven}, InfoSeek(\textit{IT2T})~\cite{chen2023infoseek}, FashionIQ(\textit{IT2I})~\cite{wu2021fashioniq}, CIRR(\textit{IT2I})~\cite{liu2021cirr} \\ \midrule
    \small$M \to M$ & OVEN(\textit{IT2IT})~\cite{hu2023oven}, InfoSeek(\textit{IT2IT})~\cite{chen2023infoseek} \\
    \bottomrule
\end{tabularx}
\end{table}

%% file: tables/appendix_additional_finetuning_design.tex
\begin{table}[t]
\small
\centering
\caption{\textbf{Additional ablation study of the subsequent turn design for fine-tuning on single-pair dataset.} `Self-reconstruction template' implies using the initial turn for the subsequent turn. Notably, simple text patterns for the special mask token yield comparable performance. `~~~~' is 3 space characters.
}
\label{tab:apdx_additional_finetuning_design}
\vspace{-0.5em}
\begin{tabular}{lc}
\toprule
\textbf{Setup} & \textbf{MMEB} \\ \midrule
Self-reconstruction template & 68.5 \\ \midrule
Simple text pattern for masking (`~~~~') & 69.4 \\
Simple text pattern for masking (`\_\_\_') & 69.4 \\
Special token for masking (\texttt{<|mask|>}) & 69.5 \\
\bottomrule
\end{tabular}
\vspace{-0.5em}
\end{table}

%% file: tables/ablation_finetuning_combination.tex
\begin{table}[t]
\footnotesize
\centering
\caption{\textbf{Various augmented data combinations (Eq. 6).} We report MMEB performance for various data combinations using \ours-2B model. Note that Case 1 represents the baseline result obtained by fine-tuning without applying the \ours\ strategy for fine-tuning.}
\label{tab:abl_finetuning_pair_combinations}
\vspace{-1em}
\rowcolors{1}{}{gray!10}
\begin{tabular}{c|cccc|c}
\toprule
Case & $(q,p)$ & $(q,p')$ & $(q',p)$ & $(q',p')$ & MMEB \\ \midrule
1 & \checkmark  & - & - & - & 68.1 \\
2 & - & \checkmark & - & - & 67.7 \\
3 & - & - & \checkmark & - & 67.9\\
4 & - & - & - & \checkmark & 68.0 \\
5 & \checkmark & \checkmark & - & - & 69.1 \\
6 & \checkmark & - & \checkmark & - &  68.9 \\
7 & \checkmark  & - & - & \checkmark &  68.9 \\
8 & - & \checkmark & \checkmark & - & 68.6 \\
9 & - & \checkmark & - & \checkmark & 68.7 \\
10 & - & - & \checkmark & \checkmark & 68.7 \\ \midrule
11 & \checkmark  & \checkmark & \checkmark & \checkmark & 69.5 \\ \bottomrule
\end{tabular}
\vspace{-1em}
\end{table}

%% file: tables/appendix_only_finetuning.tex
\begin{table*}[t]
\footnotesize
\centering
\caption{\textbf{Impact of pretraining and fine-tuning on MMEB.} We report Precision@1 (\%) results across four categories: Classification (\textbf{CLS}), Visual Question Answering (\textbf{VQA}), Retrieval (\textbf{RET}), and Visual Grounding (\textbf{GRD}). \textbf{ID} and \textbf{OOD} denote in-distribution and out-of-distribution averages, respectively. Notably, our model using \ours\ fine-tuning alone already surpasses previous state-of-the-art performance.}
\vspace{-1em}
\label{tab:apdx_only_finetuning_mmeb}
\begin{tabular}{cc|cccc|cc|c}
\toprule
\textbf{Pretraining} & \textbf{Fine-tuning} & \textbf{CLS} & \textbf{VQA} & \textbf{RET} & \textbf{GRD} & \textbf{ID} & \textbf{OOD} & \textbf{Overall} \\
\midrule
\multicolumn{9}{c}{\textbf{\ours-2B}} \\
\midrule
\checkmark & -- & 53.6 & 59.9 & 55.2 & 74.6 & -- & -- & 58.2 \\
-- & \checkmark & 62.4 & 64.5 & 70.2 & 85.1 & 72.9 & 62.2 & 68.4 \\
\checkmark & \checkmark & 66.2 & 65.6 & 70.1 & 85.8 & 72.9 & 65.0 & 69.5 \\
\midrule
\multicolumn{9}{c}{\textbf{\ours-7B}} \\
\midrule
\checkmark & -- & 56.0 & 64.7 & 58.9 & 75.7 & -- & -- & 61.6 \\
-- & \checkmark & 68.6 & 70.7 & 72.0 & 89.7 & 76.7 & 67.6 & 72.6 \\
\checkmark & \checkmark & 68.3 & 71.9 & 73.7 & 90.9 & 77.3 & 69.1 & 73.6 \\
\bottomrule
\end{tabular}
\end{table*}

\begin{table*}[t]
\footnotesize
\centering
\caption{\textbf{Impact of pretraining and fine-tuning on M-BEIR.} We report average Recall (\%) results where $S$ and $M$ denote Single modality (text or image) and Multi-modality (text and image), respectively. The arrow ($\to$) indicates the `query $\to$ target' direction.}
\vspace{-1em}
\label{tab:apdx_only_finetuning_mbeir}
\begin{tabular}{cc|cccc|c}
\toprule
\textbf{Pretraining} & \textbf{Fine-tuning} &\small $S \to S$ &\small $S \to M$ &\small $M \to S$ &\small $M \to M$ & \textbf{Overall} \\
\midrule
\multicolumn{7}{c}{\textbf{\ours-2B}} \\
\midrule
\checkmark & -- & 15.1 & 21.7 & 11.0 & 35.4 & 17.4  \\
-- & \checkmark & 48.5 & 67.3 & 41.2 & 63.5 & 50.9  \\
\checkmark & \checkmark & 49.1 & 68.2 & 41.6 & 64.8 & 51.6  \\
\midrule
\multicolumn{7}{c}{\textbf{\ours-7B}} \\
\midrule
\checkmark & -- & 16.7 & 29.1 & 11.9 & 33.7 & 19.2 \\
-- & \checkmark & 52.8 & 70.4 & 46.3 & 69.3 & 55.5  \\
\checkmark & \checkmark & 54.0 & 71.6 & 47.4 & 70.4 & 56.6  \\
\bottomrule
\end{tabular}
\vspace{-1em}
\end{table*}

%% file: tables/appendix_extended_mbeir.tex
\begin{table*}[t]
\centering
\caption{\textbf{Extended main table on M-BEIR.} The first row specifies the retrieval task, where $q$ and $c$ denote query and candidate, with superscripts $t$ and $i$ representing text and image modalities, respectively. Abbreviations include VN (VisualNews), F200K (Fashion200K), InfoS (InfoSeek), and FIQ (FashionIQ). We report Recall@10 for FashionIQ and Fashion200K, and Recall@5 for all other datasets.
} 
\label{tab:apdx_extended_mbeir}
\resizebox{\linewidth}{!}{
\begin{tabular}
{llc@{\hspace{0.1cm}}c@{\hspace{0.1cm}}c@{\hspace{0.1cm}}c@{\hspace{0.1cm}}c@{\hspace{0.1cm}}c@{\hspace{0.1cm}}c@{\hspace{0.1cm}}c@{\hspace{0.1cm}}c@{\hspace{0.1cm}}c@{\hspace{0.1cm}}c@{\hspace{0.1cm}}c@{\hspace{0.1cm}}c@{\hspace{0.1cm}}c@{\hspace{0.1cm}}c@{\hspace{0.1cm}}c@{\hspace{0.1cm}}c}

\toprule
 & & \multicolumn{3}{c}{$q^t \to c^i$} & {$q^t \to c^t$} & \multicolumn{2}{c}{{$q^t \to (c^i, c^t)$}} & \multicolumn{3}{c}{{$q^i \to c^t$}} & {$q^i \to c^i$} & \multicolumn{2}{c}{{$(q^i, q^t) \to c^t$}} & \multicolumn{2}{c}{{$(q^i, q^t) \to c^i$}} & \multicolumn{2}{c}{{$(q^i, q^t) \to (c^i, c^t)$}} & \\
 \cmidrule(r){3-5} \cmidrule(r){6-6}  \cmidrule(r){7-8} \cmidrule(r){9-11} \cmidrule(r){12-12} \cmidrule(r){13-14} \cmidrule(r){15-16} \cmidrule(r){17-18} 
 \textbf{Methods} & \textbf{Size} & \textbf{VN}  & \textbf{COCO} & \textbf{F200K} & \textbf{WebQA} & \textbf{EDIS} & \textbf{WebQA} & \textbf{VN} & \textbf{COCO} & \textbf{F200K} & \textbf{NIGHTS} & \textbf{OVEN} & \textbf{InfoS} & \textbf{FIQ} & \textbf{CIRR} & \textbf{OVEN} & \textbf{InfoS} & \textbf{Overall} \\
\cmidrule(r){3-5} \cmidrule(r){6-6}  \cmidrule(r){7-8} \cmidrule(r){9-11} \cmidrule(r){12-12} \cmidrule(r){13-14} \cmidrule(r){15-16} \cmidrule(r){17-18} 
& & \textbf{R@5} & \textbf{R@5} & \textbf{R@10} & \textbf{R@5} & \textbf{R@5} & \textbf{R@5} & \textbf{R@5} & \textbf{R@5} & \textbf{R@10} & \textbf{R@5} & \textbf{R@5} & \textbf{R@5} & \textbf{R@10} & \textbf{R@5} & \textbf{R@5} & \textbf{R@5} & \\
\midrule
\multicolumn{19}{c}{\textit{Zeroshot setting (pretrained) on M-BEIR benchmark}} \\
\midrule
CLIP-L~\cite{radford2021learning} & 428M & 0.0 & 0.0 & 0.0 & 32.1 & 6.7 & 5.5 & 0.0 & 0.0   & 0.0 & 25.3 & 0.0 & 0.0 & 4.4 & 5.4 & 24.5 & 22.1 & 9.9      \\
SigLIP~\cite{zhai2023sigmoid} & 652M & 0.0 & 0.0 & 0.0 & 34.0 & 1.1 & 2.1 & 0.0 & 0.0   & 0.0 & 28.7 & 0.0 & 0.0 & 4.8 & 7.1 & 27.2 & 24.3 & 8.1  \\
BLIP~\cite{li2022blip} & 470M & 0.0 & 0.0 & 0.0 & 38.1 & 0.0 & 0.0 & 0.0 & 0.0   & 0.0 & 25.1 & 0.0 & 0.0 & 2.2 & 7.4 & 10.1 & 7.9 & 5.7  \\
BLIP2~\cite{li2023blip2} & 210M & 0.0 & 0.0 & 0.0 & 35.2 & 0.0 & 0.0 & 0.0 & 0.0   & 0.0 & 24.0 & 0.0 & 0.0 & 3.9 & 6.2 & 13.8 & 11.4 & 5.9  \\
\midrule
\ours-2B & 2B & 3.4 & 6.7 & 0.2 & 64.4 & 28.4 & 15.0 & 1.9 & 17.6 & 0.2 & 26.0 & 7.3 & 8.6 & 10.2 & 18.1 & 38.0 & 32.9 & 17.4 \\
\ours-7B & 7B & 4.1 & 7.6 & 0.1 & 72.6 & 31.3 & 26.9 & 2.0 & 20.1 & 0.2 & 26.6 & 8.2 & 8.7 & 10.3 & 20.5 & 36.0 & 31.5 & 19.2 \\
\midrule
\multicolumn{19}{c}{\textit{Fine-tuning on M-BEIR benchmark ($<7B$ Models)}} \\
\midrule
$\text{UniIR}$~\cite{wei2024uniir} & 428M & 42.6 & 77.9 & 17.8 & 84.7 & 59.4 & 78.8 & 42.8 & 92.3 & 17.9 & 32.0 & 39.2 & 24.0 & 24.3 & 43.9 & 60.2 & 44.6 & 48.9 \\
LamRA-Ret~\cite{liu2025lamra} & 2B & 31.5 & 70.8 & 23.1 & 82.9 & 54.2 & 78.3 & 31.3 & 88.3 & 24.5 & 28.8 & 45.8 & 42.4 & 30.2 & 45.5 & 67.0 & 55.1 & 50.0 \\ \midrule
\ours-2B & 2B & 32.3 & 63.3 & 27.9 & 89.0 & 54.5 & 82.0 & 33.7 & 90.8 & 26.0 & 29.9 & 45.0 & 44.5 & 29.9 & 47.1 & 66.3 & 63.3 & 51.6 \\
\midrule
\multicolumn{19}{c}{\textit{Fine-tuning on M-BEIR benchmark ($\geq7B$ Models)}} \\
\midrule
MM-Embed~\cite{lin2025mmembed} & 7B & 41.0 & 71.3 & 17.1 & 95.9 & 68.8 & 85.0 & 41.3 & 90.1 & 18.4 & 32.4 & 42.1 & 42.3 & 25.7 & 50.0 & 64.1 & 57.7 & 52.7 \\
LamRA-Ret~\cite{liu2025lamra} & 7B & 41.3 & 75.4 & 28.7 & 85.8 & 62.5 & 81.0 & 39.3 & 90.4 & 30.4 & 32.1 & 48.4 & 48.7 & 33.1 & 50.5 & 70.0 & 60.0 & 54.9 \\
M3Task-UEM~\cite{sharma2025multi} & 7B & 40.1 & 82.1 & 30.7 & 80.5 & 67.8 & 82.0 & 44.4 & 93.4 & 31.0 & 29.8 & 51.7 & 31.9 & 31.4 & 52.5 & 71.4 & 40.0 & 53.9 \\ \midrule
\ours-7B & 7B & 41.7 & 74.0 & 32.3 & 91.8 & 59.7 & 83.6 & 39.2 & 92.3 & 31.6 & 29.5 & 51.1 & 51.7 & 33.3 & 53.4 & 72.0 & 68.8 & 56.6 \\
\bottomrule
 \end{tabular}
}
\vspace{-1em}
\end{table*}

\begin{table*}[t]
\centering
\caption{\textbf{Extended main table on M-BEIR$_\text{local}$.} This table presents quantitative results evaluated using \textbf{local candidate pools} specific to each of the 16 datasets. The first row specifies the retrieval task, where $q$ and $c$ denote query and candidate, with superscripts $t$ and $i$ representing text and image modalities, respectively. Abbreviations include VN (VisualNews), F200K (Fashion200K), InfoS (InfoSeek), and FIQ (FashionIQ). We report Recall@10 for FashionIQ and Fashion200K, and Recall@5 for all other datasets.
} 
\label{tab:apdx_extended_mbeir_local}
\resizebox{\linewidth}{!}{
\begin{tabular}
{llc@{\hspace{0.1cm}}c@{\hspace{0.1cm}}c@{\hspace{0.1cm}}c@{\hspace{0.1cm}}c@{\hspace{0.1cm}}c@{\hspace{0.1cm}}c@{\hspace{0.1cm}}c@{\hspace{0.1cm}}c@{\hspace{0.1cm}}c@{\hspace{0.1cm}}c@{\hspace{0.1cm}}c@{\hspace{0.1cm}}c@{\hspace{0.1cm}}c@{\hspace{0.1cm}}c@{\hspace{0.1cm}}c@{\hspace{0.1cm}}c}

\toprule
 & & \multicolumn{3}{c}{$q^t \to c^i$} & {$q^t \to c^t$} & \multicolumn{2}{c}{{$q^t \to (c^i, c^t)$}} & \multicolumn{3}{c}{{$q^i \to c^t$}} & {$q^i \to c^i$} & \multicolumn{2}{c}{{$(q^i, q^t) \to c^t$}} & \multicolumn{2}{c}{{$(q^i, q^t) \to c^i$}} & \multicolumn{2}{c}{{$(q^i, q^t) \to (c^i, c^t)$}} & \\
 \cmidrule(r){3-5} \cmidrule(r){6-6}  \cmidrule(r){7-8} \cmidrule(r){9-11} \cmidrule(r){12-12} \cmidrule(r){13-14} \cmidrule(r){15-16} \cmidrule(r){17-18} 
 \textbf{Methods} & \textbf{Size} & \textbf{VN}  & \textbf{COCO} & \textbf{F200K} & \textbf{WebQA} & \textbf{EDIS} & \textbf{WebQA} & \textbf{VN} & \textbf{COCO} & \textbf{F200K} & \textbf{NIGHTS} & \textbf{OVEN} & \textbf{InfoS} & \textbf{FIQ} & \textbf{CIRR} & \textbf{OVEN} & \textbf{InfoS} & \textbf{Overall} \\
\cmidrule(r){3-5} \cmidrule(r){6-6}  \cmidrule(r){7-8} \cmidrule(r){9-11} \cmidrule(r){12-12} \cmidrule(r){13-14} \cmidrule(r){15-16} \cmidrule(r){17-18} 
& & \textbf{R@5} & \textbf{R@5} & \textbf{R@10} & \textbf{R@5} & \textbf{R@5} & \textbf{R@5} & \textbf{R@5} & \textbf{R@5} & \textbf{R@10} & \textbf{R@5} & \textbf{R@5} & \textbf{R@5} & \textbf{R@10} & \textbf{R@5} & \textbf{R@5} & \textbf{R@5} & \\
\midrule
\multicolumn{19}{c}{\textit{Zeroshot setting (pretrained) on M-BEIR benchmark}} \\
\midrule
CLIP-L~\cite{radford2021learning} & 428M & 43.3 & 61.1 & 6.6 & 36.2 & 43.3 & 45.1 & 41.3 & 79.0   & 7.7 & 26.1 & 24.2 & 20.5 & 7.0 & 13.2 & 38.8 & 26.4 & 32.5      \\
SigLIP~\cite{zhai2023sigmoid} & 652M & 30.1 & 75.7 & 36.5 & 39.8 & 27.0 & 43.5 & 30.8 & 88.2   & 34.2 & 28.9 & 29.7 & 25.1 & 14.4 & 22.7 & 41.7 & 27.4 & 37.2  \\
BLIP~\cite{li2022blip} & 470M & 16.4 & 74.4 & 15.9 & 44.9 & 26.8 & 20.3 & 17.2 & 83.2   & 19.9 & 27.4 & 16.1 & 10.2 & 2.3 & 10.6 & 27.4 & 16.6 & 26.8  \\
BLIP2~\cite{li2023blip2} & 210M & 16.7 & 63.8 & 14.0 & 38.6 & 26.9 & 24.5 & 15.0 & 80.0   & 14.2 & 25.4 & 12.2 & 5.5 & 4.4 & 11.8 & 27.38 & 15.8 & 24.8  \\
\midrule
\ours-2B & 2B & 18.5 & 66.0 & 13.8 & 71.0 & 46.0 & 49.0 & 17.8 & 79.4 & 10.9 & 28.0 & 31.6 & 39.3 & 9.8 & 27.4 & 48.2 & 48.3 & 37.8 \\
\ours-7B & 7B & 18.8 & 68.6 & 15.6 & 79.1 & 49.7 & 58.2 & 18.5 & 80.9 & 13.0 & 29.7 & 37.6 & 41.1 & 12.8 & 29.1 & 49.5 & 49.1 & 40.7 \\
\midrule
\multicolumn{19}{c}{\textit{Fine-tuning on M-BEIR benchmark ($<7B$ Models)}} \\
\midrule
$\text{UniIR}$~\cite{wei2024uniir} & 428M & 42.6 & 81.1 & 18.0 & 84.7 & 59.4 & 78.7 & 43.1 & 92.3 & 18.3 & 32.0 & 45.5 & 27.9 & 24.4 & 44.6 & 67.6 & 48.9 & 50.6 \\
LamRA-Ret~\cite{liu2025lamra} & 2B & 30.8 & 79.7 & 25.1 & 80.8 & 54.3 & 77.8 & 31.2 & 88.9 & 27.1 & 28.7 & 51.1 & 44.2 & 28.9 & 47.7 & 72.3 & 60.8 & 51.8 \\ \midrule
\ours-2B & 2B & 33.2 & 80.3 & 28.2 & 89.5 & 55.3 & 81.3 & 32.8 & 90.4 & 26.8 & 30.0 & 51.9 & 48.0 & 30.7 & 51.4 & 72.2 & 65.2 & 54.2 \\
\midrule
\multicolumn{19}{c}{\textit{Fine-tuning on M-BEIR benchmark ($\geq7B$ Models)}} \\
\midrule
LamRA-Ret~\cite{liu2025lamra} & 7B & 41.6 & 81.5 & 28.7 & 86.0 & 62.6 & 81.2 & 39.6 & 90.6 & 30.4 & 30.4 & 32.1 & 54.1 & 52.1 & 33.1 & 76.2 & 63.3 & 56.6 \\
\midrule
\ours-7B & 7B & 42.2 & 83.3 & 33.2 & 90.9 & 60.2 & 83.1 & 40.6 & 92.5 & 32.2 & 28.6 & 57.3 & 55.8 & 33.6 & 56.7 & 76.2 & 72.3 & 58.7 \\
\bottomrule
 \end{tabular}
}
\vspace{-1em}
\end{table*}

%% file: tables/appendix_extended_mmeb.tex
\begin{table*}[t]
\centering
\caption{\textbf{Extended main table on MMEB.} This table details the results for both baselines and \ours\ on the MMEB benchmark. The evaluation covers 20 in-distribution and 16 out-of-distribution datasets, where out-of-distribution entries are distinguished by a yellow background. We specifically feature our strongest model \ours-7B}
\resizebox{\linewidth}{!}{
\begin{tabular}{lcccccccccccc}
\toprule
& \multicolumn{4}{c}{\textbf{Zeroshot}} & \multicolumn{8}{c}{\textbf{Fine-tuning}} \\
\cmidrule(lr){2-5} \cmidrule(lr){6-13}
& \textbf{CLIP} & \textbf{MMRet} & \textbf{mmE5} & \textbf{\ours} & \textbf{VLM2VEC} & \textbf{MMRet} & \textbf{mmE5} & \textbf{LLaVE} & \textbf{UniME} & \textbf{B3} & \textbf{MoCa} & \textbf{\ours}\\
\midrule

\rowcolor{orange!30} \textbf{Classification (10 tasks)} & & & & & & & & & & & &\\
ImageNet-1K                            & 55.8 & 49.1 & 68.8 & 62.7 & 74.5 & 58.8 & 77.8 & 77.1 & 71.3 & 84.3 & 78.0 & 82.9 \\
N24News                                & 34.7 & 45.8 & 54.5 & 48.1 & 80.3 & 71.3 & 81.7 & 82.1 & 79.5 & 81.6 & 81.5 & 82.0 \\
HatefulMemes                           & 51.1 & 51.0 & 55.0 & 58.9 & 67.9 & 53.7 & 64.2 & 74.3 & 64.6 & 64.2 & 77.6 & 73.4 \\
VOC2007                                & 50.7 & 74.6 & 73.9 & 67.1 & 91.5 & 85.0 & 91.0 & 90.3 & 90.4 & 89.7 & 90.0 & 86.5 \\
SUN397                                 & 43.4 & 60.1 & 72.7 & 73.1 & 75.8 & 70.0 & 77.7 & 79.1 & 75.9 & 82.8 & 76.8 & 80.7 \\
\rowcolor{yellow!15} Place365          & 28.5 & 35.3 & 39.7 & 40.8 & 44.0 & 43.0 & 43.0 & 45.0 & 45.6 & 47.9 & 43.0 & 46.0 \\
\rowcolor{yellow!15} ImageNet-A        & 25.5 & 31.6 & 46.1 & 24.0 & 43.6 & 36.1 & 56.3 & 51.6 & 45.5 & 56.5 & 52.7 & 61.8 \\
\rowcolor{yellow!15} ImageNet-R        & 75.6 & 66.2 & 86.2 & 89.5 & 79.8 & 71.6 & 86.3 & 90.9 & 78.4 & 91.9 & 83.0 & 91.9 \\
\rowcolor{yellow!15} ObjectNet         & 43.4 & 49.2 & 74.8 & 70.9 & 39.6 & 55.8 & 62.5 & 46.2 & 36.4 & 73.2 & 45.2 & 49.9 \\
\rowcolor{yellow!15} Country-211       & 19.2 &  9.3 & 35.1 & 25.2 & 14.7 & 14.7 & 35.4 & 20.1 & 18.7 & 27.8 & 30.4 & 28.3 \\
\textit{All Classification}            & 42.8 & 47.2 & 60.7 & 62.6 & 61.2 & 56.0 & 67.6 & 65.7 & 60.6 & 70.0 & 65.8 & 68.3 \\
\midrule

\rowcolor{blue!30} \textbf{VQA (10 tasks)} & & & & & & & & & & & &\\
OK-VQA                                 & 7.5  & 28.0 & 56.6 & 62.6 & 69.0 & 73.3 & 67.6 & 71.1 & 68.3 & 71.6 & 36.9 & 72.7 \\
A-OKVQA                                & 3.8  & 11.6 & 50.0 & 53.4 & 54.4 & 56.7 & 56.1 & 70.8 & 58.7 & 59.5 & 57.1 & 60.4 \\
DocVQA                                 & 4.0  & 12.6 & 81.3 & 92.0 & 52.0 & 78.5 & 90.3 & 90.3 & 67.6 & 94.7 & 94.3 & 95.4 \\
InfographicsVQA                        & 4.6  & 10.6 & 44.0 & 68.6 & 30.7 & 39.3 & 56.5 & 53.5 & 37.0 & 68.9 & 77.2 & 78.0 \\
ChartQA                                & 1.4  & 2.4  & 35.2 & 50.6 & 34.8 & 41.7 & 50.5 & 62.2 & 33.4 & 59.8 & 69.8 & 68.2 \\
Visual7W                               & 4.0  & 9.0  & 40.4 & 48.4 & 49.8 & 49.5 & 51.9 & 55.8 & 51.7 & 55.9 & 58.5 & 64.1 \\
\rowcolor{yellow!15} ScienceQA         & 9.4  & 23.3 & 47.3 & 52.9 & 42.1 & 45.2 & 55.8 & 54.4 & 40.5 & 51.7 & 59.2 & 57.3 \\
\rowcolor{yellow!15} VizWiz            & 8.2  & 25.9 & 54.0 & 50.1 & 43.0 & 51.7 & 52.8 & 48.5 & 42.7 & 50.6 & 46.2 & 55.0 \\
\rowcolor{yellow!15} GQA               & 41.3 & 41.3 & 65.4 & 84.4 & 61.2 & 59.0 & 61.7 & 68.4 & 63.6 & 67.5 & 71.6 & 81.4 \\
\rowcolor{yellow!15} TextVQA           & 7.0  & 18.9 & 83.1 & 84.4 & 62.0 & 79.0 & 83.3 & 79.4 & 65.2 & 85.1 & 75.8 & 86.4 \\
\textit{All VQA}                       & 9.1  & 18.4 & 55.7 & 65.0 & 49.9 & 57.4 & 62.6 & 65.4 & 52.9 & 66.5 & 64.7 & 71.9 \\
\midrule

\rowcolor{green!30} \textbf{Retrieval (12 tasks)} & & & & & & & & & & & &\\
VisDial                                & 30.7 & 62.6 & 39.1 & 65.0 & 80.9 & 83.0 & 74.1 & 83.0 & 79.7 & 86.1 & 84.5 & 85.3 \\
CIRR                                   & 12.6 & 65.7 & 41.6 & 27.6 & 49.9 & 61.4 & 54.7 & 54.5 & 52.2 & 65.8 & 53.4 & 54.1 \\
VisualNews\_t2i                        & 78.9 & 45.7 & 51.2 & 48.7 & 75.4 & 74.2 & 77.6 & 76.6 & 74.8 & 80.7 & 78.2 & 82.9 \\
VisualNews\_i2t                        & 79.6 & 33.4 & 64.9 & 60.4 & 80.0 & 78.1 & 83.3 & 81.2 & 78.8 & 84.5 & 83.1 & 85.6 \\
MSCOCO\_t2i                            & 59.5 & 68.7 & 55.0 & 67.5 & 75.7 & 78.6 & 76.4 & 78.9 & 74.9 & 79.8 & 79.8 & 79.3 \\
MSCOCO\_i2t                            & 57.7 & 56.7 & 59.1 & 62.3 & 73.1 & 72.4 & 73.2 & 74.7 & 73.8 & 76.7 & 73.9 & 77.2 \\
NIGHTS                                 & 60.4 & 59.4 & 58.9 & 64.6 & 65.5 & 68.3 & 68.3 & 67.0 & 66.2 & 67.4 & 66.7 & 66.6 \\
WebQA                                  & 67.5 & 76.3 & 82.9 & 83.7 & 87.6 & 90.2 & 88.0 & 90.4 & 89.8 & 90.4 & 91.4 & 89.2 \\
\rowcolor{yellow!15} FashionIQ         & 11.4 & 31.5 & 21.6 & 17.9 & 16.2 & 54.9 & 28.8 & 23.3 & 16.5 & 28.2 & 28.9 & 24.7 \\
\rowcolor{yellow!15} Wiki-SS-NQ        & 55.0 & 25.4 & 58.8 & 70.6 & 60.2 & 24.9 & 65.8 & 63.9 & 66.6 & 69.5 & 82.7 & 73.5 \\
\rowcolor{yellow!15} OVEN              & 41.1 & 73.0 & 67.6 & 62.3 & 56.5 & 87.5 & 77.5 & 68.0 & 55.7 & 70.6 & 80.4 & 74.5 \\
\rowcolor{yellow!15} EDIS              & 81.0 & 59.9 & 55.2 & 76.7 & 87.8 & 65.6 & 83.7 & 89.1 & 86.2 & 88.7 & 96.9 & 92.0 \\
\textit{All Retrieval}                 & 53.0 & 56.5 & 54.7 & 58.9 & 67.4 & 69.9 & 71.0 & 70.9 & 67.9 & 74.1 & 75.0 & 73.7 \\
\midrule

\rowcolor{purple!30} \textbf{Visual Grounding (4 tasks)} & & & & & & & & & & & &\\
MSCOCO                                 & 33.8 & 42.7 & 59.0 & 53.6 & 80.6 & 76.8 & 53.7 & 87.0 & 76.5 & 74.4 & 84.6 & 80.6 \\
\rowcolor{yellow!15} RefCOCO           & 56.9 & 69.3 & 78.9 & 79.3 & 88.7 & 89.8 & 92.7 & 95.4 & 89.3 & 92.9 & 94.0 & 94.2 \\
\rowcolor{yellow!15} RefCOCO-matching  & 61.3 & 63.2 & 80.8 & 91.5 & 84.0 & 90.6 & 88.8 & 92.8 & 90.6 & 91.2 & 95.5 & 93.6 \\
\rowcolor{yellow!15} Visual7W-pointing & 55.1 & 73.5 & 71.2 & 78.5 & 90.9 & 77.0 & 92.3 & 92.5 & 84.1 & 80.5 & 95.3 & 95.1 \\
\textit{All Visual Grounding}          & 51.8 & 62.2 & 72.5 & 75.7 & 86.1 & 83.6 & 89.6 & 91.9 & 85.1 & 84.6 & 92.4 & 90.9 \\
\midrule

\rowcolor{cyan!15} \textbf{Final Score (36 tasks)} & & & & & & & & & & & &\\
All IND                                & 37.1 & 43.5 & 57.2 & 60.9 & 67.5 & 59.1 & 72.3 & 64.4 & 68.4 & 75.9 & 74.7 & 77.3 \\
All OOD                                & 38.7 & 44.3 & 60.4 & 62.4 & 57.1 & 68.0 & 66.7 & 75.0 & 57.9 & 67.1 & 67.6 & 69.1 \\
All                                    & 37.8 & 44.0 & 58.6 & 61.6 & 62.9 & 64.1 & 69.8 & 70.3 & 66.6 & 72.0 & 71.5 & 73.6 \\
\bottomrule
\end{tabular}
}
\label{tab:apdx_extended_mmeb}
\end{table*}